\documentclass{aa}  
\usepackage{courier,graphicx,epsfig,color,colordvi,natbib,url,twoopt,ulem}
\usepackage[varg]{txfonts}
\pdfoutput=1
\usepackage{hyperref} 
\usepackage{pdfcomment,acronym}

\bibpunct{(}{)}{;}{a}{}{,}   

\def\# #1\par{\par\mbox{}\\ \noindent{\color{red}\small $\sharp$ #1}\\} 

\bibpunct{(}{)}{;}{a}{}{,}    
\makeatletter
 \newcommandtwoopt{\citeads}[3][][]{%
   \nonstopmode
   \href{http://adsabs.harvard.edu/abs/#3}%
        {\def\hyper@linkstart##1##2{}%
         \let\hyper@linkend\@empty\citealp[#1][#2]{#3}}
   \biblink{#3}{\href{http://adsabs.harvard.edu/abs/#3}{ADS}}%
   \errorstopmode}            
 \newcommandtwoopt{\citepads}[3][][]{%
   \nonstopmode
   \href{http://adsabs.harvard.edu/abs/#3}%
        {\def\hyper@linkstart##1##2{}%
         \let\hyper@linkend\@empty\citep[#1][#2]{#3}}
   \biblink{#3}{\href{http://adsabs.harvard.edu/abs/#3}{ADS}}%
   \errorstopmode}            
 \newcommandtwoopt{\citetads}[3][][]{%
   \nonstopmode
   \href{http://adsabs.harvard.edu/abs/#3}%
        {\def\hyper@linkstart##1##2{}%
         \let\hyper@linkend\@empty\citet[#1][#2]{#3}}
   \biblink{#3}{\href{http://adsabs.harvard.edu/abs/#3}{ADS}}%
   \errorstopmode}            
 \newcommandtwoopt{\citeyearads}[3][][]{%
   \nonstopmode
   \href{http://adsabs.harvard.edu/abs/#3}%
        {\def\hyper@linkstart##1##2{}%
         \let\hyper@linkend\@empty\citeyear[#1][#2]{#3}}
   \biblink{#3}{\href{http://adsabs.harvard.edu/abs/#3}{ADS}}%
   \errorstopmode}            
\makeatother

\newcount\longrefs
\def\aap{\ifnum\longrefs=1 {Astron.\ Astrophys.}\else 
                           {A\hbox{\rm \&}A}\fi}
\def\aapr{\ifnum\longrefs=1 {Astron.\ Astrophys.\ Rev.}\else 
                            {A\hbox{\rm \&}AR}\fi}
\def\aaps{\ifnum\longrefs=1 {Astron.\ Astrophys.\ Suppl.}\else 
                            {A\hbox{\rm \&}A Suppl.}\fi}
\def\actaa{\ifnum\longrefs=1 {Acta Astronomica}\else
                            {Acta Astron.}\fi}
\def\aipcs{\ifnum\longrefs=1 {Am.\ Inst.\ Phys.\ Conf.\ Series}\else
                             {AIP Conf.\ Ser.}\fi}
\def\aj{\ifnum\longrefs=1 {Astron.\ J.}\else 
                          {AJ}\fi} 
\def\ao{\ifnum\longrefs=1 {Applied Optics}\else 
                           {Appl.\ Opt.}\fi} 
\def\aspcs{\ifnum\longrefs=1 {Astron.\ Soc.\ Pacific Conf.\ Series}\else 
                           {ASP Conf.\ Ser.}\fi} 
\def\apj{\ifnum\longrefs=1 {Astrophys.\ J.}\else 
                           {ApJ}\fi} 
\def\apjl{\ifnum\longrefs=1 {Astrophys.\ J. Lett.}\else 
                            {ApJL}\fi} 
\def\aplett{\ifnum\longrefs=1 {Astrophys.\ J. Lett.}\else 
                            {ApJ}\fi} 
\def\apjs{\ifnum\longrefs=1 {Astrophys.\ J. Suppl.}\else 
                            {ApJS}\fi}
\def\apss{\ifnum\longrefs=1 {Astrophys.\ and Space Science}\else 
                            {Astrophys.\ Space Sci.}\fi}
\def\araa{\ifnum\longrefs=1 {Ann.\ Rev.\ Astron.\ Astrophys.}\else 
                            {ARA\hbox{\rm \&}A}\fi}
\def\azh{\ifnum\longrefs=1 {Astronomicheskii Zhurnal}\else 
                            {Astron.\ Zhur.}\fi}
\def\baas{\ifnum\longrefs=1 {Bull.\ Am.\ Astron.\ Soc.}\else 
                            {BAAS}\fi}
\def\bain{\ifnum\longrefs=1 {Bull.\ Astronom.\ Institutes Netherlands}\else
                            {Bull.\ Astr.\ Inst.\ Neth.}\fi}
\def\cjaa{\ifnum\longrefs=1 {Chinese Jour.\ Astron.\ Astrophys.}\else 
                            {Chin.\ J.\ A\&A}\fi}
\def\gca{\ifnum\longrefs=1 {Geochim.\ Cosmochim.\ Acta}\else 
                           {Geochim.\ Cosmochim.\ Acta}\fi}
\def\grl{\ifnum\longrefs=1 {Geophys.\ Res.\ Lett.}\else 
                           {Geoph.\ Res.\ Lett.}\fi}
\def\iaucirc{\ifnum\longrefs=1 {IAU Circulars}\else 
                          {IAU Circ.}\fi}
\def\icarus{\ifnum\longrefs=1 {Icarus}\else 
                          {Icarus}\fi}
\def\ip{\ifnum\longrefs=1 {in press}\else 
                          {in press}\fi}
\def\jcap{\ifnum\longrefs=1 {Jour.\ Cosmology Astropart.\ Phys.}\else 
                          {JCAP}\fi}
\def\jgr{\ifnum\longrefs=1 {J.\ Geophys.\ Res.}\else 
                           {J.\ Geophys.\ Res.}\fi}  
\def\jrasc{\ifnum\longrefs=1 {J.\ Royal Astron.\ Soc.\ Canada}\else 
                           {JRAS Can.}\fi}  
\def\memsai{\ifnum\longrefs=1 {Mem.~Soc.~Astron.~Italiana}\else
                              {MmSAI}\fi}
\def\mnras{\ifnum\longrefs=1 {Mon.\ Not.\ Roy.\ Astron.\ Soc.}\else 
                             {MNRAS}\fi} 
\def\na{\ifnum\longrefs=1 {New Astronomy}\else 
                           {New Astron.}\fi}
\def\nar{\ifnum\longrefs=1 {New Astronomy rev.}\else 
                           {New Astron.\ Rev.}\fi}
\def\nat{\ifnum\longrefs=1 {Nature}\else 
                           {Nat}\fi}
\def\pasa{\ifnum\longrefs=1 {Pub.\ Astron.\ Soc.\ Australia}\else 
                            {PASA}\fi} 
\def\pasj{\ifnum\longrefs=1 {Pub.\ Astron.\ Soc.\ Japan}\else 
                            {PASJ}\fi} 
\def\pasp{\ifnum\longrefs=1 {Pub.\ Astron.\ Soc.\ Pacific}\else 
                            {PASP}\fi} 
\def\physscr{\ifnum\longrefs=1 {Physica Scripta}\else 
                            {Phys.\ Scrip.}\fi} 
\def\planss{\ifnum\longrefs=1 {Planetary \& Space Science}\else 
                            {Plan. \& Space Sci.}\fi} 
\def\procspie{\ifnum\longrefs=1 {Proc.\ SPIE}\else 
                            {Proc.\ SPIE}\fi} 
\def\qjras{\ifnum\longrefs=1 {Quarterly J.\ Royal Astron.\ Soc.}\else 
                            {QJRAS}\fi} 
\def\rmxaa{\ifnum\longrefs=1 {Revista Mexicana de Astron.\ y Astrofys.}\else 
                            {RMxAA}\fi} 
\def\sa{\ifnum\longrefs=1 {Soviet Astron..}\else 
                               {Sov.\ Astron.}\fi}
\def\skytel{\ifnum\longrefs=1 {Sky \& Telescope}\else 
                            {Sky \& Tel.}\fi} 
\def\solphys{\ifnum\longrefs=1 {Solar Phys.}\else 
                               {SoPh}\fi}
\def\sovast{\ifnum\longrefs=1 {Soviet Astronomy}\else 
                               {Sov.\ Ast.}\fi}
\def\ssr{\ifnum\longrefs=1 {Space Science Rev.}\else 
                               {Space\ Sci.\ Rev.}\fi}
\def\zap{\ifnum\longrefs=1 {Zeitschr.\ f.\ Astrophysik}\else
                               {Z.\ Astrophys.}\fi}

\makeatletter
\newcommand{\bibnote}[2]{\@namedef{#1note}{#2}}
\newcommand{\biblink}[2]{\@namedef{#1link}{#2}}
\makeatother

\newacro{AA}{Astronomy \& Astrophysics}
\newacro{ADS}{Astrophysics Data System}
\newacro{AIA}{Atmospheric Imaging Assembly}
\newacro{AO}{adaptive optics}
\newacro{ApJ}{Astrophysical Journal}
\newacro{AR}{active region}
\newacro{BFI}{Broad-band Filter Imager}
\newacro{CE}{coronal equilibrium}
\newacro{CfA}{Center for Astrophysics}
\newacro{CME}{coronal mass ejection}
\newacro{CRD}{complete redistribution}
\newacro{CRISP}{CRisp Imaging SpectroPolarimeter}
\newacro{CRISPEX}{CRisp SPectral EXplorer}
\newacro{CS}{coherent scattering}
\newacro{DEM}{Differential Emission Measure}
\newacro{DF}{dynamic fibril}
\newacro{DKIST}{Daniel K. Inouye Solar Telescope}
\newacro{DLR}{Deutsches Zentrum f\"ur Luft- und Raumfahrt}
\newacro{DOT}{Dutch Open Telescope}
\newacro{DST}{Richard B. Dunn Solar Telescope}   
\newacro{EB}{Ellerman bomb}
\newacro{EDP}{\'{E}dition Diffusion Presse Sciences}  
\newacro{EIT}{Extreme ultraviolet Imaging Telescope}
\newacro{EPIC}{European participation in Solar-C}
\newacro{ERC}{European Research Council}
\newacro{ESA}{European Space Agency}
\newacro{EST}{European Solar Telescope}
\newacro{EUV}{extreme ultraviolet}
\newacro{FAF}{flaring active-region fibril}
\newacro{FITS}{Flexible Image Transport System}
\newacro{FOV}{field of view}
\newacro{fov}{field of view}
\newacro{FWHM}{full width at half maximum}
\newacro{HAO}{High Altitude Observatory}
\newacro{HD}{hydrodynamics}
\newacro{Hi-C}{High Resolution Coronal Imager Sounding Rocket}
\newacro{HMI}{Helioseismic and Magnetic Imager}
\newacro{IAA}{Instituto de Astrof\'{i}sica de Andaluc\'{i}a}
\newacro{IAC}{Instituto de Astrof\'{i}sica de Canarias}
\newacro{IAS}{Institut d'Astrophysique Spatiale}
\newacro{IDL}{Interactive Data Language}
\newacro{IMaX}{Imaging Magnetograph eXperiment}
\newacro{INAF}{Istituto Nazionale di Astrofisica}
\newacro{IB}{IRIS bomb}
\newacro{IR}{infrared}
\newacro{IRIS}{Interface Region Imaging Spectrograph}
\newacro{ISAS}{Institute of Space and Astronautical Science}
\newacro{ISP}{Institute for Solar Physics}
\newacro{ISS}{International Space Station}
\newacro{ISSI}{International Space Science Institute}
\newacro{ITA}{Institute for Theoretical Astrophysics}
\newacro{JAXA}{Japan Aerospace Exploration Agency}
\newacro{KIS}{Kiepenheuer--Institut f\"{u}r Sonnenphysik}
\newacro{KPNO}{Kitt Peak National Observatory}
\newacro{LASP}{Laboratory for Atmospheric and Space Physics}
\newacro{LC}{liquid cristal}
\newacro{LMSAL}{Lockheed Martin Solar and Astrophysics Labratory}
\newacro{LOS}{line of sight}
\newacro{LTE}{local thermodynamic equilibrium}
\newacro{MC}{magnetic concentration}
\newacro{MCAO}{multi-conjugate adaptive optics} 
\newacro{MDI}{Michelson Doppler Imager}
\newacro{ME}{Milne-Eddington} 
\newacro{MHD}{magnetohydrodynamics}
\newacro{MOMFBD}{Multi-Object Multi-Frame Blind Deconvolution}
\newacro{MPE}{Max--Planck--Institut f\"ur extraterrestrische Physik}
\newacro{MPG}{Max--Planck--Gesellschaft}
\newacro{MPS}{Max Planck Institute for Solar System Research}
\newacro{MSSL}{Mullard Space Science Laboratory}
\newacro{MTF}{modulation transfer function}
\newacro{NAOJ}{National Astronomical Observatory of Japan}
\newacro{NASA}{National Aeronautics and Space Administration}
\newacro{NLTE}{non-local thermodynamic equilibrium}
\newacro{NLFFF}{non-linear force-free field}
\newacro{NOAA}{National Oceanic and Atmospheric Administration}
\newacro{non-E}{non-equilibrium}
\newacro{NSO}{National Solar Observatory}
\newacro{NWO}{Netherlands Organisation for Scientific Research}
\newacro{PRD}{partial redistribution}
\newacro{PROBA2}{PRoject for OnBoard Autonomy}
\newacro{PSF}{point spread function}
\newacro{QS}{quiet Sun}
\newacro{QSEB}{quiet-Sun Ellerman-like brightening} 
\newacro{RAL}{Rutherford Appleton Laboratory}
\newacro{RBE}{rapid blue-shifted excursion}
\newacro{R-MHD}{radiation hydrodynamics}
\newacro{rms}{root mean square}
\newacro{RMS}{root mean square}
\newacro{ROB}{Royal Observatory of Belgium}
\newacro{ROI}{region of interest}
\newacro{RRE}{rapid red-shifted excursion}
\newacro{RTE}{radiative transfer equation}
\newacro{SE}{statistical equilibrium}
\newacro{SB}{Saha Boltzmann}
\newacro{SDO}{Solar Dynamics Observatory}
\newacro{SJI}{slit-jaw image}
\newacro{SNR}{signal-to-noise ratio}
\newacro{SO}{Solar Orbiter}
\newacro{SoHO}{Solar and Heliospheric Observatory}
\newacro{SP}{Spectropolarimeter}
\newacro{SST}{Swedish 1-m Solar Telescope}
\newacro{SUMER}{Solar Ultraviolet Measurements of Emitted Radiation}
\newacro{SUFI}{Sunrise Filter Imager}
\newacro{SVD}{singular value decomposition}
\newacro{SVST}{Swedish Vacuum Solar Telescope}
\newacro{THEMIS}{T\'{e}lescope H\'{e}liographique pour l'Etude du 
   Magn\'{e}tisme et des Instabilit\'{e} Solaires}     
\newacro{TR}{transition region}
\newacro{TRACE}{Transition Region and Coronal Explorer}
\newacro{TSI}{total solar irradiance}
\newacro{UT}{Universal Time}
\newacro{UV}{ultraviolet}
\newacro{VAULT}{Very high angular resolution ultraviolet telescope}
\newacro{VIRGO}{Variability of solar IRradiance and Gravity Oscillations}
\newacro{VTT}{Vacuum Tower Telescope}    
\newacro{XRT}{X-Ray Telescope}

\long\def\startignore #1\stopignore{}   

\def\ie{\rmit{i.e.,}}              
\def\eg{\rmit{e.g.,}}              

\def\specchar#1{\uppercase{#1}}    
\def\specand{ and }                
\def\specand{\,\&\,}               

\def\CII{\mbox{C\,\specchar{ii}}}

\def\CaII{\mbox{Ca\,\specchar{ii}}}

\def\FeI{\mbox{Fe\,\specchar{i}}} 
\def\FeII{\mbox{Fe\,\specchar{ii}}}

\def\HeII{\mbox{He\,\specchar{ii}}}

\def\MgII{\mbox{Mg\,\specchar{ii}}}

\def\NiI{\mbox{Ni\,\specchar{i}}}

\def\OI{\mbox{O\,\specchar{i}}}

\def\SiIV{\mbox{Si\,\specchar{iv}}}

\def\Halpha{\mbox{H\hspace{0.1ex}$\alpha$}} 

\def\Hbeta{\mbox{H\hspace{0.2ex}$\beta$}}

\def\HeIDthree{\mbox{He\,\specchar{i}\,\,D$_{3}$}}

\def\CaIIK{\mbox{Ca\,\specchar{ii}\,\,K}}       
\def\CaIIH{\mbox{Ca\,\specchar{ii}\,\,H}}
\def\CaIIHK{\mbox{Ca\,\specchar{ii}\,\,H{\specand}K}}

\def\Kthree{\mbox{K$_3$}}      

\def\Ktwo{\mbox{K$_2$}}

\def\KtwoV{\mbox{K$_{2V}$}}
\def\KtwoR{\mbox{K$_{2R}$}}

\def\CaIR{\mbox{Ca\,\specchar{ii}\,8542\,\AA}} 

\def\MgIIk{\mbox{Mg\,\specchar{ii}\,\,k}}
\def\MgIIh{\mbox{Mg\,\specchar{ii}\,\,h}}
\def\MgIIhk{\mbox{Mg\,\specchar{ii}{\specand}k}}

\def\kthree{\mbox{k$_3$}}    

\def\ktwo{\mbox{k$_2$}}

\def\ktwoV{\mbox{k$_{2V}$}}
\def\ktwoR{\mbox{k$_{2R}$}}

\def\level #1 #2#3#4{$#1 \; ^{#2} \mbox{#3} ^{#4}$}   

\def\deg{\hbox{$^\circ$}}       

\def\arcsec{\hbox{$^{\prime\prime}$}}

\def\kms{\hbox{km\;s$^{-1}$}}

\def\tis{\!=\!}                             
\def\={\hbox{$\!=\!$}}                     

\def\specchar#1{{\sc{#1}}}    

\def\dakK{\mbox{$\times10^{4}$\;K}}
\def\cgsint{\hbox{erg\;s$^{-1}$\;cm$^{-2}$\;Hz$^{-1}$\;sr$^{-1}$}}
\def\vlos{\hbox{$v_{\rm{LOS}}$}}
\def\vturb{\hbox{$v_{\rm{micro}}$}}
\def\Blon{\hbox{$B_{\rm{lon}}$}}
\def\Bhor{\hbox{$B_{\rm{hor}}$}}
\def\ltau{\hbox{log $\tau_{500}$}}
\def\dT{\hbox{$\Delta T$}}
\def\EBs{Ellerman bombs}
\def\EB{Ellerman bomb}
\def\UVBs{UV bursts}
\def\UVB{UV burst}
\def\CRISP{CRisp Imaging SpectroPolarimeter}
\def\SST{Swedish 1-m Solar Telescope}
\def\IRIS{{\it Interface Region Imaging Spectrograph\/}}
\def\STiC{STockholm Inversion Code}
\def\Hinode{{\it Hinode\/}}
\def\AIA{{\it Atmospheric Imaging Assembly\/}}
\def\SDO{{\it Solar Dynamics Observatory\/}}
\def\EBDETECT{{\tt EBDETECT}}
\def\Bifrost{{\it Bifrost\/}}
\renewcommand{\ie}{i.e.}
\renewcommand{\eg}{e.g.}

\begin{document}

\title{Dissecting bombs and bursts: non-LTE inversions of low-atmosphere
reconnection in SST and IRIS observations}
 
\author{G.~J.~M.~Vissers$^{1}$ 
\and J.~de la Cruz Rodr{\'{\i}}guez$^{1}$  
\and T.~Libbrecht$^{1}$
\and L.~H.~M.~Rouppe van der Voort$^{2,3}$
\and G.~B.~Scharmer$^{1}$
\and M.~Carlsson$^{2,3}$
}
\institute{Institute for Solar Physics, Department of Astronomy, 
 Stockholm University, AlbaNova University Centre,
 106 91 Stockholm, Sweden
\and Institute of Theoretical Astrophysics,
  University of Oslo, %
P.O. Box 1029 Blindern, N-0315 Oslo, Norway
\and Rosseland Centre for Solar Physics, 
  University of Oslo, %
P.O. Box 1029 Blindern, N-0315 Oslo, Norway
}

\titlerunning{Non-LTE inversions of low-atmosphere reconnection}
\authorrunning{G.~J.~M.~Vissers et al.}

\date{}

\abstract{\EBs\ and \UVBs\ are transient brightenings that are ubiquitously observed in
  the lower atmospheres of active and emerging flux regions. 
  As they are believed to pinpoint sites of magnetic reconnection in
  reconfiguring fields, understanding their occurrence and detailed evolution
  may provide useful insights in the overall evolution of active regions. 
  Here we present results from inversions of SST/CRISP and CHROMIS, as well as
  IRIS data of such transient events. 
  Combining information from the \MgIIhk, \SiIV\ and \CaIR\ and \CaIIHK\ lines,
  we aim to characterise their temperature and velocity stratification, as well
  as their magnetic field configuration.
  We find average temperature enhancements of a few thousand kelvin close
  to the classical temperature minimum, similar to previous studies, but
  localised peak temperatures of up to 10,000--15,000\,K from \CaII\ inversions.
  Including \MgII\ appears to generally dampen these temperature enhancements to
  below 8000\,K, while \SiIV\ requires temperatures in
  excess of 10,000\,K at low heights, but may also be reproduced with
  secondary temperature enhancements of 35,000--60,000\,K higher up.
  However, reproducing \SiIV\ comes at the expense of overestimating the \MgII\
  emission.
  The line-of-sight velocity maps show clear bi-directional jet signatures for
  some events and strong correlation with substructure in the intensity images
  in general.
  Absolute line-of-sight velocities range between 5--20\,\kms\ on average, with
  slightly larger velocities towards the observer than away.
  The inverted magnetic field parameters show an enhancement of the horizontal
  field co-located with the brightenings at heights similar to that of the
  temperature increase.
  We are thus able to largely reproduce the observational properties of \EBs\
  with \UVB\ signature (\eg\ intensities, profile asymmetries, morphology and
  bi-directional jet signatures) with temperature stratifications peaking close
  to the classical temperature minimum.
  Correctly modelling the \SiIV\ emission in agreement with all other
  diagnostics is, however, an outstanding issue and remains paramount in
  explaining its apparent coincidence with \Halpha\ emission.
  Fine-tuning the approach (accounting for resolution differences, fitting localised
  temperature enhancements and/or performing spatially-coupled inversions) is
  likely necessary in order to obtain better agreement between all considered
  diagnostics.
} 

\maketitle

\section{Introduction}\label{sec:introduction}
Active and emerging flux regions host an abundance of compact transient
brightenings, particularly in their early evolving stages, that are the
likely heating signatures of reconnection as emerging fields reconfigure.
Examples of such transients range from \EBs\ in the lower atmosphere to
microflares in the upper chromosphere and understanding their formation may
therefore provide essential information in understanding the evolution of active
regions as a whole.

First observed in 1915, \EBs\ were described in a publication two years later
\citepads{1917ApJ....46..298E}  
and have been subject of renewed interest since the early 2000s with the
observations from the Flare Genesis balloon mission
\citepads{2002ApJ...575..506G}, 
but in particular since the high-resolution imaging 
with the Solar Optical Telescope (SOT) aboard \Hinode\ and imaging spectroscopy
with the \CRISP\ (CRISP;
\citeads{2008ApJ...689L..69S}) 
at the
\SST\ (SST;
\citeads{2003SPIE.4853..341S}). 
Both \Hinode\ observations in \CaIIH\ 
\citepads{2010PASJ...62..879H} 
and CRISP observations in \Halpha\ 
\citepads{2011ApJ...736...71W} 
clearly demonstrated \EB\ sub-arcsecond fine-structure and rapid
variability on a timescale of seconds.
Furthermore, imaging spectroscopy with the SST revealed that these are
sub-canopy events: while they are clearly visible in the wings of \Halpha\ they
get more and more obscured as one observes closer to line centre, to the point
that they become invisible in line core images (%
\citeads{2011ApJ...736...71W}, 
\citeads{2013JPhCS.440a2007R}). 

A phenomenon with similar morphology and dynamics was identified in early
\IRIS\ (IRIS; 
\citeads{2014SoPh..289.2733D}) 
observations by
\citetads{2014Sci...346C.315P} 
and described as ``hot bombs'' which were suggested to be located in the cool
lower atmosphere.
These \UVBs\ are characterised by strongly broadened and enhanced \SiIV,
\CII\ and \MgIIhk\ lines, often with absorption blends from neutral species
superimposed. 
While the latter already suggests sub-canopy formation of the emission, these
events share further characteristics with \EBs: they tend to occur on polarity
inversion lines, have a signature in the UV continua at 1600\,\AA\ and
1700\,\AA\ observed by the \SDO's (SDO) 
\AIA\ (AIA;
\citeads{2012SoPh..275...17L}), 
but remain invisible in its
\HeII\ and higher-temperature coronal channels.
Without co-temporal \Halpha\ data a connection to \EBs\ could not be made at the
time, but later studies (\eg\
\citeads{2015ApJ...812...11V}, 
\mbox{\citeads{2015ApJ...810...38K}}, 
\citeads{2016ApJ...824...96T}, 
\citeads{2017A&A...598A..33L}),  
have shown that there is indeed overlap between the \EB\ and \UVB\ populations,
however not one-to-one.

Recent 3D magneto-hydrodynamic numerical experiments have reproduced the typical
\Halpha\ wing enhancements observed in Quiet Sun \EB-like events 
(\citeads{2017A&A...601A.122D};  
these are the Quiet Sun counterparts of the ``classical'' \EBs\ and were first
reported by 
\citeads{2016A&A...592A.100R}) 
and in stronger-field \EBs\
\citepads{2017ApJ...839...22H}. 
The latter study was also able to reproduce the \SiIV\ enhancements that
characterise \UVBs, albeit not simultaneously with the \EB\ signatures; 
\ie\ the events with \EB\ signature did not show enhanced \SiIV\ emission,
while the \UVBs\ had enhanced \Halpha\ core intensity, unlike observational
\EBs.
The reconnection height appears to be key: where \EBs\ resulted from
reconnection in the first few hundred kilometers of the atmosphere, \UVBs\ were
due to reconnection up at some 2\,Mm.
This may help explain the observational characteristic that not all \EBs\ have a
\UVB\ counterpart signature (cf.~\eg\
\citeads{2015ApJ...812...11V}, 
\citeads{2016ApJ...824...96T}, 
\citeads{2016A&A...593A..32G}), 
as the absence of a one-to-one correlation suggests differences in the
atmospheric conditions between events that show either signature in isolation.
It does, however, not explain those events where \SiIV\ and \Halpha\ appear
co-spatially even at more slanted lines-of-sight.

The \SiIV\ visibility poses additional problems, as this would seem to require
excessive temperatures in the lower solar atmosphere compared to what has so far
been suggested based on semi-empirical modeling of \Halpha\ and \CaII\
diagnostics (\eg\
\citeads{1983SoPh...87..135K}, 
\citeads{2010MmSAI..81..646B}, 
\citeads{2013A&A...557A.102B}, 
\citeads{2014A&A...567A.110B}, 
\citeads{2017RAA....17...31F}), 
and more recently including IRIS \MgIIh\
\citepads{2016A&A...593A..32G}. 
On the other hand, analysis of \HeIDthree\ observations with the TRIPPEL
spectrograph at the SST suggests temperatures of order a few ten thousand kelvin
could be reached
\citepads{2017A&A...598A..33L}.  
Furthermore, 
\citetads{2016A&A...590A.124R} 
argues that temperatures of order 1--2\,\dakK\ may be sufficient to result in \SiIV\
emission, provided one assumes LTE in the \EB\ onset and non-equilibrium
conditions in the subsequent dynamical evolution.

Now, comissioning observations with CHROMIS in \CaIIK\ uncover a whole new level
of fine structure, with highly dynamic blob-like substructure evolving on the
time scale of seconds.
In a recent paper, 
\citetads{2017ApJ...851L...6R} 
argue that these observations suggest plasmoid-driven reconnection in \UVBs.
This appears to be supported by 2.5D numerical experiments, where the
superposition of plasmoids at different Doppler shifts could explain
multi-peaked and triangular \SiIV\ profiles that are sometimes observed in
\UVBs.

This study aims at inferring the atmospheric stratification of \EBs\ with \UVB\
signature by combing the wealth of information that the SST and IRIS provide.
The remainder of this paper is structured as follows. 
Section~\ref{sec:observations} details the IRIS and SST observations,
including the alignment procedure and event identification and selection. 
Section~\ref{sec:stic} describes the inversion code and setup, while the
inversion results are presented in Section~\ref{sec:results}.
Section~\ref{sec:discussion} offers a discussion of these results and, finally,
in section~\ref{sec:conclusion} we summarise our conclusions.

\begin{figure*}[ht]
  \centerline{\includegraphics[width=\textwidth]{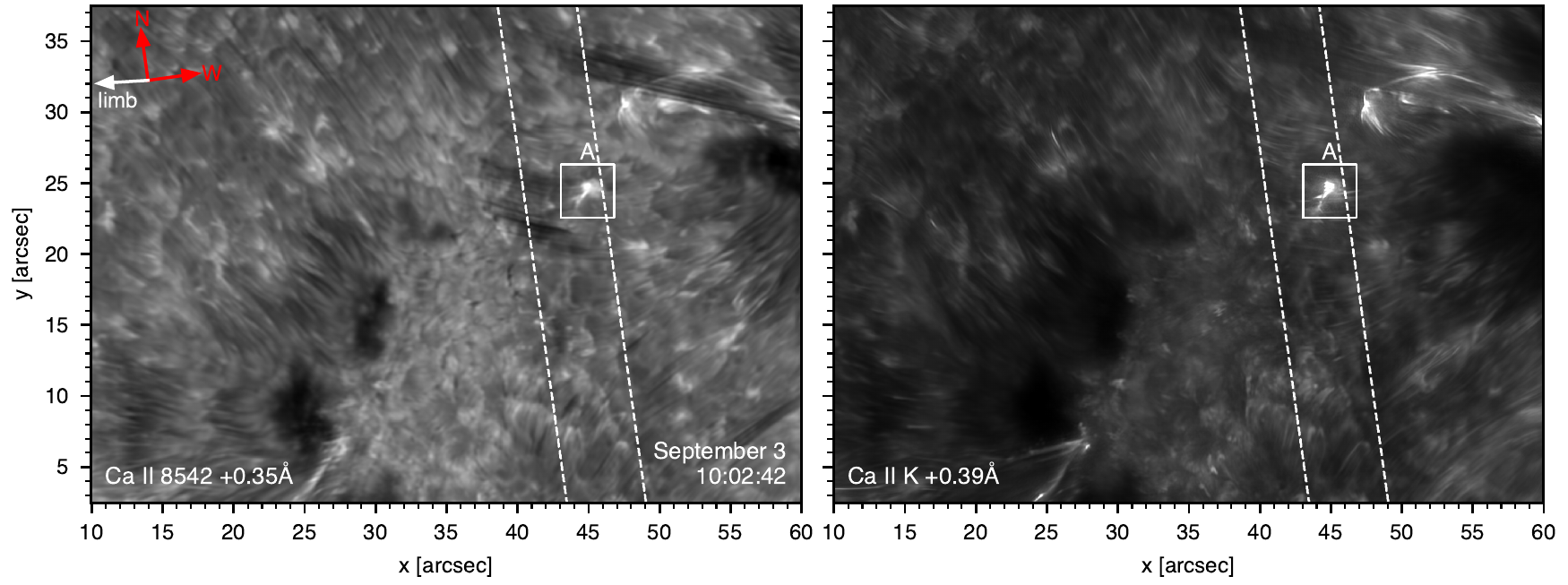}}
  \vspace{-5ex}
  \centerline{\includegraphics[width=\textwidth]{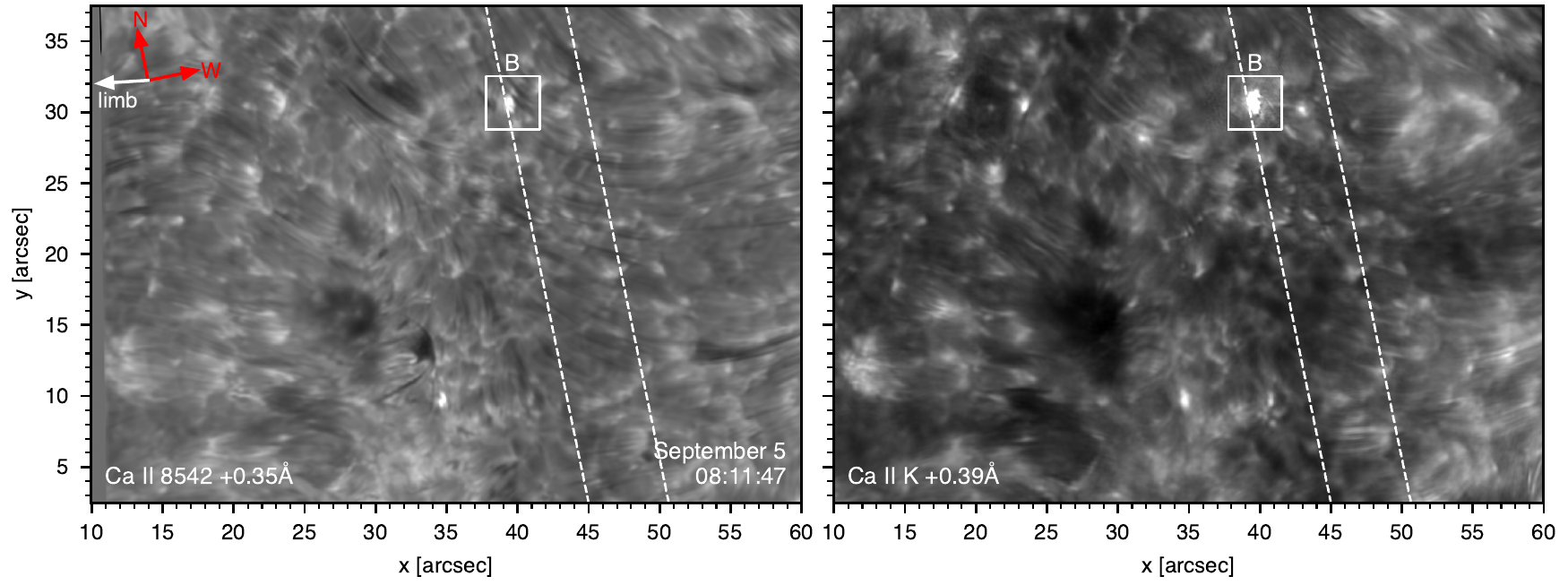}}
  \vspace{-2ex}
  \caption[]{\label{fig:fov} %
    Overview images of the 3 and 5 September, 2016 data sets in the red wing of
    \CaIR\ at +0.35\,\AA\ ({\it left panels\/}) and \CaIIK\ at +0.39\,\AA\ ({\it
    right panels\/}). 
    The location of events selected for further study are highlighted with
    labelled white boxes in the panels.
    The slanted dashed lines indicate the extent of the IRIS raster (which
    extends beyond the upper and lower boundaries of this field-of-view). 
    The red arrows point to Solar North and West, while the white arrows
    indicate the direction to the closest limb.
  }
\end{figure*}

\section{Observations and data reduction}\label{sec:observations}
\subsection{Acquisition and data properties}
For this study we employ two data sets obtained on September 3 and 5,
2016, respectively. 
On both days the target was active region NOAA 12585, with the SST pointing at
($X$,$Y$)=($-$561\arcsec,44\arcsec) on the 3rd and at
($X$,$Y$)=($-$161\arcsec,24\arcsec) on the
5th, corresponding to a viewing angle of $\mu$=0.81 and 0.99, respectively.

At the SST observations were performed with the CRISP and CHROMIS instruments.
Both are dual Fabry-P\'erot tunable filter instruments, where CRISP has 
additional polarimetric capabilities.
On both days, CRISP provided imaging spectroscopy in the \Halpha\ line in 15
positions out to $\pm$1.5\,\AA\ at 200\,m\AA\ steps and full Stokes
imaging spectropolarimetry in the \CaIR\ line in 21 positions out to
$\pm$1.75\,\AA\ at 70\,m\AA\ steps in the inner wings and increasing steps of up
to 800\,m\AA\ in the outer wings.  The cadence of these observations is 20\,s.
On September 5 this sequence was extended to include full Stokes
spectropolarimetry in the \FeI~6301 and 6302\,\AA\ lines in 16 wavelength
positions, resulting in an overall cadence of 32\,s.
On both days CHROMIS recorded \CaIIK\ and \Hbeta\ imaging spectroscopy, but for
our analysis we focus only on the former.
The \CaIIK\ line was sampled out to $\pm$0.7\,\AA\ at 78\,m\AA\ spacing and
additional samplings at $\pm$1.33\,\AA\ as well as a continuum point out at
3999.7\,\AA.
The cadence of the CHROMIS data is 13\,s and 12\,s for the respective data sets.

On both days these observations were supported by IRIS with a medium dense
16-step raster (OBSID 3625503135).
This program covers about 5\arcsec\,$\times$\,60\arcsec\ with continuous
0\farcs{33} steps at 0.5\,s exposure time per slit position, resulting in an
overall raster cadence of 20.8\,s. 
As part of the program, the far-UV (FUV) data were spectrally rebinned
on-board by 4 to increase the signal-to-noise ratio.
Context slit-jaw imaging in the \CII, \SiIV, and \MgIIk\ bands was recorded at
10.4\,s cadence.

Absolute wavelength and intensity calibrations were performed for all data.
For CRISP and CHROMIS data the atlas profile by 
\citetads{1984SoPh...90..205N}  
was used, taking into account limb darkening due to the non-vertical viewing
angles.
For IRIS spectra we followed the standard procedure, with wavelength calibration
to the \OI~1355.5977\,\AA\ line for FUV1 (containing \CII), to
\FeII~1392.817\,\AA\ for FUV2 (containing the \SiIV\ lines) and to the
\NiI~2799.474\,\AA\ line in the near-UV (NUV), while using the
wavelength-dependent response functions for the intensity calibration.
All resulting intensities are expressed in CGS units as function of frequency
(\ie\ $I_{\nu}$ [\cgsint]).

\subsection{Data reduction and alignment}
The CRISP data were reduced using the CRISPRED
\citepads{2015A&A...573A..40D} 
processing pipeline, which includes image reconstruction through Multi-Object
Multi-Frame Blind Deconvolution (MOMFBD; 
\citeads{2005SoPh..228..191V}). 
CHROMIS data were reduced using similar procedures, modified from CRISPRED to
accomodate for the CHROMIS data format and bundled into the CHROMISRED pipeline
\citepads{2018arXiv180403030L}. 

The CRISP data were then scaled up to the native CHROMIS pixel scale (from
0\farcs{0592} to 0\farcs{0376}) and subsequently aligned to the CHROMIS images
by iteratively cross-correlating a wavelength-integrated image for every time
step in \CaIIK\ ($\pm$0.47\,\AA\ around the core) with the nearest-neighbour one
in time in \CaIR\ ($\pm$0.45\,\AA\ around the core).
Similarly, the IRIS to SST alignment was performed using the \MgIIk\ slit-jaw
images (also scaled up to CHROMIS pixel size), with wavelength-integrated \CaII\
images as anchor. 
After initial guess alignment based on pointing coordinates and FOV rotation,
further fine-alignment was achieved through iterative shift and
cross-correlation of the images until the correction shift fell below 0.1
(CHROMIS) pixel.

\subsection{Event identification and selection}
We used the output from an \EB\ detection code 
\EBDETECT\ 
\citepads{2019arXiv190107975V} 
as a basis for event selection, as comparison with the intensity images then
allows us to identify those \EBs\ that also display \UVB\ signatures.
CRISPEX (%
\citeads{2012ApJ...750...22V};  
\citeads{2018arXiv180403030L}) 
was used for data browsing, event and snapshot selection, as well as
verification of the automated detection.

Ideally, we would select events that show both \EB\ and \UVB\ signatures, as
well as events that show only one of those characteristics in isolation, but
unfortunately the data at hand only provided examples of the former.
Although comparison of the SST and IRIS slit-jaw image fields-of-view indicate a
number of events was observed that only had one of the signatures, these were
not covered by the IRIS raster.
Hence, we selected two events---A and B---for detailed study, observed on
September 3 and 5, respectively.

The spectral criterion for an event to classify as a \UVB\ is to display
profiles as described in
\citetads{2018SSRv..214..120Y}, 
\ie\ substantially enhanced and broadened \SiIV\ lines, although 
the \CII\ and \MgIIhk\ lines are commonly also enhanced.
An \EB\ with \UVB\ signature requires the same IRIS line enhancements in
addition to the regular \EB\ signature of bright \Halpha\ wings and dark core. 
Both the events under scrutiny have previously been analysed in
\citetads{2017ApJ...851L...6R}. 

For both events \CaIR, \CaIIK\ and IRIS data are available, while Event B was
also covered by additional \FeI\ spectropolarimetry.
The results we present and discuss in the following sections are from inversions
of selected snapshots of both events, as well as (temporally downsampled)
time sequence of Event A. Before presenting the inversion results, we first
discuss the inversion code and setup in the following Section~\ref{sec:stic}.

\section{Inversions with the \STiC} \label{sec:stic}
We use the MPI-parallel non-LTE \STiC\ (STiC;
\citeads{2016ApJ...830L..30D}, 
\citeads{2019A&A...623A..74D}) 
to invert the SST and IRIS line profiles in order to infer the possible
atmospheric conditions.
The code builds on an optimised version of RH 
\citepads{2001ApJ...557..389U}  
to solve the atom population densities
and in each iteration the pressure scale is
computed assuming hydrostatic equilibrium, from which in turn the hydrogen and
electron densities are derived using an LTE equation of state (from 
\citeads{2017A&A...597A..16P}). 
The electron densities can also be derived assuming non-LTE hydrogen ionisation,
by iteratively solving the statistical equilibrium equations while imposing
charge conservation (similar to
\citeads{2007A&A...473..625L}). 
We found, however, that the latter did not significantly change our inversion
results (we refer the reader to Appendix~\ref{sec:appendix} for a results
comparison between the two approaches) and therefore decided to assume LTE
electron densities instead, with the added benefit of faster and more stable
inversions.

The inversions are performed pixel-by-pixel, \ie\ assuming 1.5D plane-parallel
atmospheres.
This means that 3D radiative transfer effects, which are important for \CaII\
line cores
(\citeads{2009ApJ...694L.128L}, 
\citeads{2009ASPC..415...87L}, 
\citeads{2018A&A...612A..28L}) 
and \MgII\ lines
(\citeads{2013ApJ...772...89L}, 
\citeads{2013ApJ...772...90L}, 
\citeads{2015ApJ...806...14P}) 
cannot be taken into account by the code, however, these should not affect the
line wings as much, where the emission of interest is observed.
Also for \SiIV\ (which is generally formed under optically thin conditions) this
is likely a minor effect.
STiC does allow including partial frequency redistribution (PRD) effects by
scattered photons
\citepads{2012A&A...543A.109L}. 

We initialise the model atmosphere from FAL-C by interpolating to 44 depth points. 
While this is unlikely to be close to the solar burst atmospheres that we are
interested in, the inversion code already roughly approaches the final results
after the first cycle. 
Modification of the initial atmosphere by moving the transition region to lower
heights or by raising the chromospheric temperature plateau did not
significantly affect the inversion outcomes.

The inversions are run in multiple cycles, with the general approach being to
use fewer nodes in the first cycle to obtain the large scale trends and more
nodes in the subsequent cycles to get a more detailed atmospheric structure (as
suggested by 
\citeads{1992ApJ...398..375R}). 
In between the cycles the model atmosphere is smoothed horizontally using a
Gaussian with a 2\,pixel full-width-at-half-maximum, to decrease the
effects of pixels where inversions failed.
This smoothing is applied at twice the node resolution, \ie\ at every node point
as well as at a point equidistantly between each node, followed by
interpolation to all other depth points.
The smoothed atmosphere is then used as input atmosphere for the subsequent
cycle.

While inverting the \CaII\ data in isolation (or even with \FeI) is a
straightforward and relatively quick process, obtaining reasonable fits when
including UV lines is non-trivial as we show in the
following.
Instead of attempting direct inversion of all available diagnostics, we first
perform two cycles with \CaII\ (and for September 5 also \FeI) data, using the
output atmosphere thereof as input atmosphere for the inversions including IRIS
diagnostics.

Table~\ref{tab:cycles} summarises the number of nodes used in each cycle for
temperature $T$, line-of-sight velocity \vlos, microturbulence (or
non-thermal velocity) \vturb, the longitudinal and horizontal
components of the magnetic field \Blon\ and \Bhor\ (in the frame of the
observer, \ie\ \Blon\ is the line-of-sight component, while \Bhor\ is that in
the plane-of-the-sky),
and azimuth $\alpha$.
The nodes are by default distributed equidistantly between $\ltau\tis0.1$ and $-$8.
The exception to this is when we include \FeI, in which case the nodes for
\Blon\ and \Bhor\ are placed at specific locations:
$\ltau\tis[-0.5,-2.0,-5.0]$ and $\ltau\tis[-0.5,-5.0]$, respectively.
The third cycle applies only for runs that include IRIS data, \ie\ cycles 1 and
2 are run with \CaII\ (and if available \FeI) only and the output atmosphere
from that second cycle is used as input atmosphere for the third cycle when
\MgII\ and \SiIV\ are included.
In principle, including more diagnostics formed at different heights would warrant
increasing the number of velocity nodes, however, tests showed this did not
generally yield better fits, hence we retained the number of velocity nodes from
the second cycle in those following.

STiC offers four choices in depth interpolation of the parameters: linear,
quadratic and cubic Bezier splines
\citepads{2013ApJ...764...33D},
and discontinuous
\citepads{2016A&A...586A..42S}. 
Tests with single pixels and small patches indicated best results were obtained
with linear interpolation when considering only SST data, but that allowing for
discontinuities was necessary to better fit \MgII. 
When including \SiIV\ best results were again obtained with linear depth
interpolation.

The model atoms used for \CaII, \MgII\ and \SiIV\ have 6, 11 and 9 levels,
respectively.
\CaIIK\ and \MgIIhk\ are computed with PRD,
while for \CaIR\ and \SiIV\ complete frequecy redistribution (CRD) is assumed.

\begin{table}[h]
\caption{Number of nodes used in each inversion cycle.}
\begin{center}
  
\begin{tabular}{l|cc|cc|cc}%
	\hline \hline
  Parameter  & \multicolumn{2}{c|}{\CaIR}  & \multicolumn{2}{c|}{\CaII\ (+\FeI)} &
  \multicolumn{2}{c}{\CaII\ (+\FeI) + IRIS} \\
  \cline{2-3}\cline{4-5}\cline{6-7} & 1 & 2 & 1 & 2 & 3A & 3B \\
	\hline
  $T$               &  4  & 9 & 4 & 9 & 9 & 13 \\
  $v_{\rm{LOS}}$    &  1  & 3 & 1 & 4 & 4 & 4  \\
  $v_{\rm{micro}}$  &  0  & 2 & 1 & 5 & 5 & 5  \\
  $B_{\rm{lon}}$    &  1  & 2 & 1 & 2 (3)$^{a}$ & 2 (3)$^{a}$ & 2  \\
  $B_{\rm{hor}}$    &  1  & 2 & 1 & 2 & 2 & 2  \\
  $\alpha$          &  1  & 1 & 1 & 1 & 1 & 1  \\
	\hline\hline
\end{tabular}
\tablefoot{
    The first two sets of columns specify the nodes for the two inversion cycles
    considering only SST spectral diagnostics, while the last two provide
    the details for the third cycle that includes IRIS diagnostics,  
    differentiating between the run adding only \MgII\ (3A) and the one including
  both \MgII\ and \SiIV\ (3B) on top of all other diagnostics. \\
  \tablefoottext{a}{The number of nodes for \Blon\ in cycles 2 and 3 depends on
  whether \FeI\ is included or not; three nodes are used when it is. Both
  \Blon\ and \Bhor\ nodes are in that case placed at particular values of \ltau\
  (see main text).} 
}
\end{center}
\label{tab:cycles}
\end{table}


\begin{figure*}[bht]
  \centerline{\includegraphics[width=\textwidth]{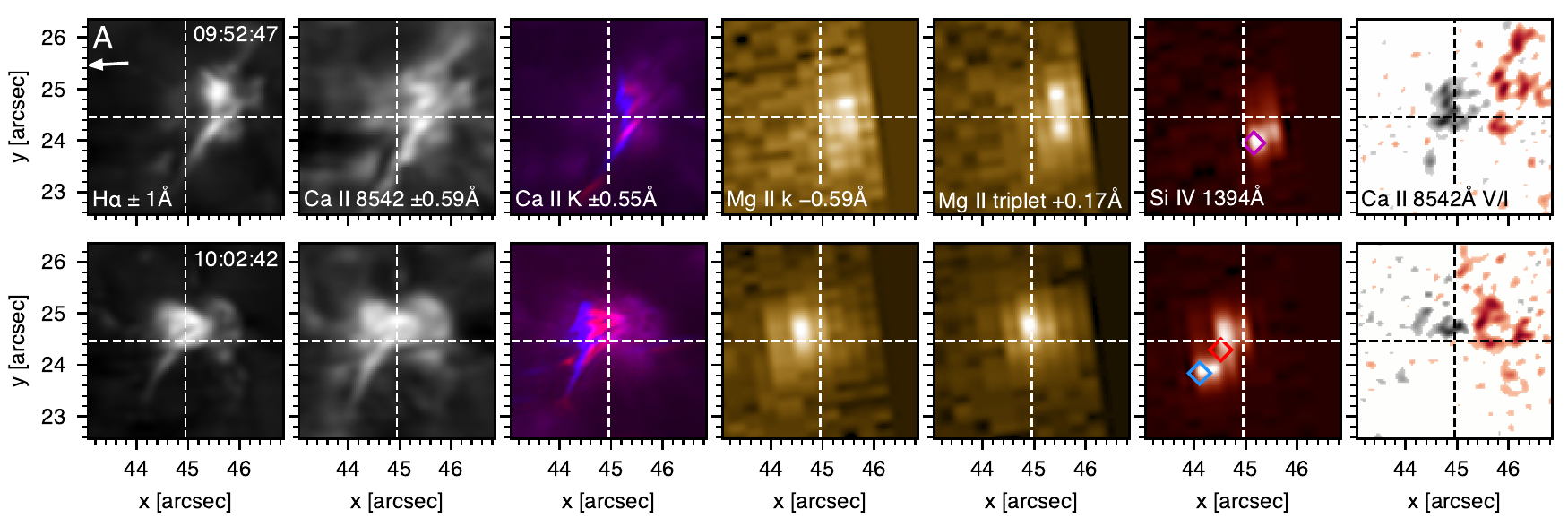}}
  \vspace{-3.5ex}
  \centerline{\includegraphics[width=\textwidth]{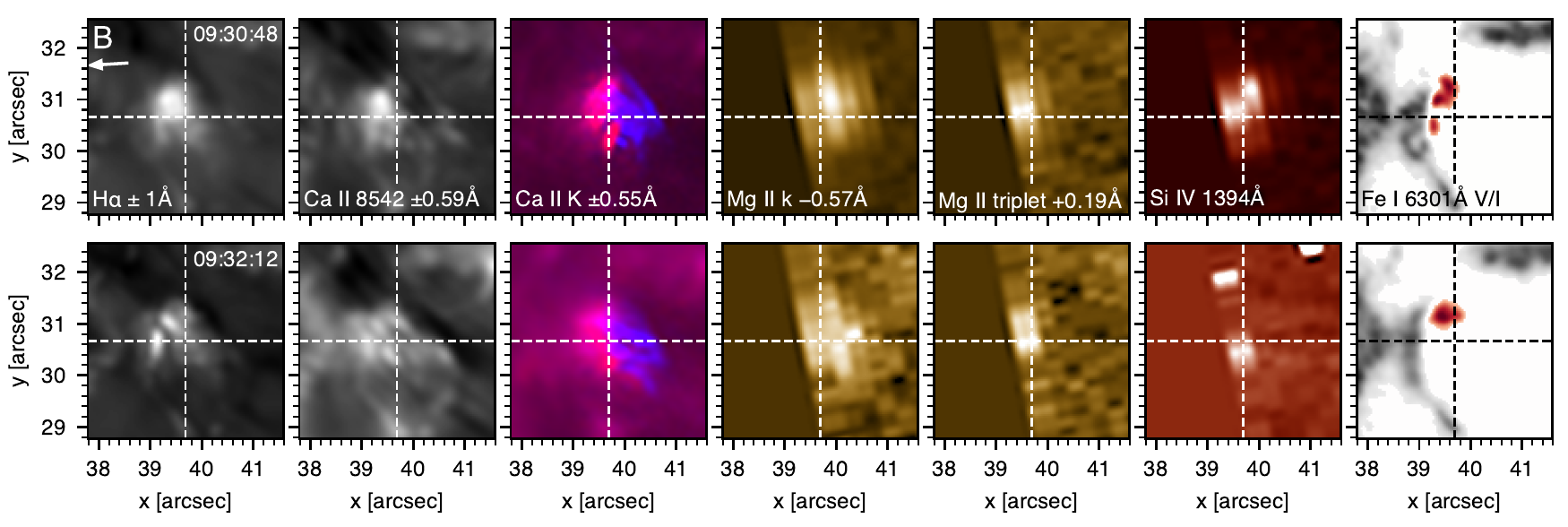}}
  \vspace{-2ex}
  \caption[]{\label{fig:subfov} %
    Close-up images of the selected events in the various diagnostics in which
    they are observed.
    {\it From left to right\/}:
    \Halpha\ summed wings ($\pm$1\,\AA), \CaIR\ summed wings
    ($\pm$0.59\,\AA), composite \CaIIK\ images at
    the indicated wavelength offsets (with blue wing in blue, red wing in red), 
    \MgIIk\ blue wing, \MgII\ triplet red wing, \SiIV\ (nominal rest wavelength)
    and a Stokes $V/I$ magnetogram proxy (from \CaIR\ for September 3 and
    \FeI\,6301.5\,\AA\ at $-$0.07\,\AA\ for September 5), with positive and negative polarities
    in black and red, respectively.
    For each event---labelled in the top left corner of the \Halpha\ panel---two
    snapshots are shown at the times specified in the top right corner of the
    same panel.
    The white arrows below the event labels indicate the direction to the
    closest limb.
    The dashed cross-hairs do not highlight any particular feature, but are
    meant to aid in comparing the substructure in various diagnostics.
    The coloured diamonds in the \SiIV\ panels of Event A (second-to-last
    panels in the top two rows) indicate locations for which spectra and
  inversion results are shown in Fig.~\ref{fig:siiv_invprof}.
    Panels have been bytescaled individually to better highlight relevant
    substructure.
  }
\end{figure*}

\section{Results} \label{sec:results}
Figure~\ref{fig:subfov} shows two snapshots of Event A in various diagnostics
in its top two rows; the next two rows offer a similar display for Event B.
The panels are selected through nearest neighbour interpolation in time with
\CaIIK\ as leading diagnostic, resulting in a difference of up to 2.9 and 9.2\,s
between the panels for the two snapshots of Event A and up to 2.3 and 5.3\,s for
those of Event B.
Depending on the frames chosen these timing differences can be much
larger, though: for CRISP and CHROMIS the scan-averaged times can differ by
nearly 20\,s and 6\,s for September 3 and 5, respectively, while the offset with
IRIS can run up to about 11\,s on both days. 
This can have an appreciable impact on the ability of STiC to obtain agreement
between the different diagnostics, in particular for those cases where
fast-evolving fine structure is considered (cf.~\eg\ 
\citetads{2018A&A...614A..73F},  
although the effects on the Stokes profiles are not as extreme in our case).
The frames displayed here were chosen for their relatively high contrast and the
amount of substructure visible in \CaIIK\ images.

For Event A \Halpha\ and \CaIR\ wing images show much the same morphology,
although the \CaIR\ wing emission is stronger in the top parts of the \EB.
Comparing the \CaIR\ and \CaIIK\ panels---and as already noted in 
\citetads{2017ApJ...851L...6R}---\CaIIK\ 
reveals additional substructure compared to the CRISP images.
Particularly the second snapshot, at 10:02\,UT, shows that the more or less
monolithic \Halpha\ and \CaIR\ brightening at the geometric base of the \EB\
(\ie\ crossed by the vertical dashed line at
$(x,y)\simeq(45\arcsec,25\arcsec)$) is composed of at least three
individual, thin jet-like structures.
Striking is also the spatial offset between the location of red and blue \Ktwo\
peak emission (third column), where in both snapshots the \KtwoR\
enhancement is located at the \EB\ base, while the \KtwoV\ enhancement is
predominantly observed as the jet tops (cf. the offset between red and
blue).
Considering the next three panels showing IRIS \MgII\ and \SiIV\ this event
classifies as an \EB\ with \UVB\ signature.
The fine structure so well-observed with CHROMIS is unsurprisingly lost, but
confirms earlier reports (\eg\
\citeads{2015ApJ...812...11V}) 
indicating the \MgIIk\ wing emission appears co-spatial with the body of the
\Halpha, while \SiIV\ is offset with respect to the bulk of both \Halpha\ and
\MgII\ emission and is mostly observed towards the geometric top of the event.
Finally, comparison with the \CaIR\ Stokes $V/I$ panel suggests opposite
polarity reconnection as the driver of this event; guided by the cross-hairs we
can readily trace the base of the \Halpha\ brightenings to the polarity
inversion line.

A similar picture emerges for Event B, although geometric effects separating
both the bi-directional red- and blue-shifts and the emission at the different
IRIS wavelengths are much smaller, as one would expect given the near-vertical
viewing angle.
Individual jets that can be seen in particularly the second \Halpha\ snapshot
are not that well visible in the \CaIR\ and \CaIIK\ panels, however both \CaIIK\
panels show clearly the presence of blob-shaped substructure that the study by 
\citetads{2017ApJ...851L...6R} 
suggested to be signatures of plasmoids.
The \MgII\ enhancements overlap largely with the \Halpha\ and \CaII\ emission,
while \SiIV\ appears somewhat more offset, however this is difficult to
establish conclusively given that the raster does not cover the entire event as
observed with the SST.
The underlying magnetogram shown in the right-most panels, in this case derived
from the blue wing of \FeI\,6301.5\,\AA, again points to opposite polarity
reconnection.

\begin{figure*}[bht]
  \centerline{\includegraphics[width=\textwidth]{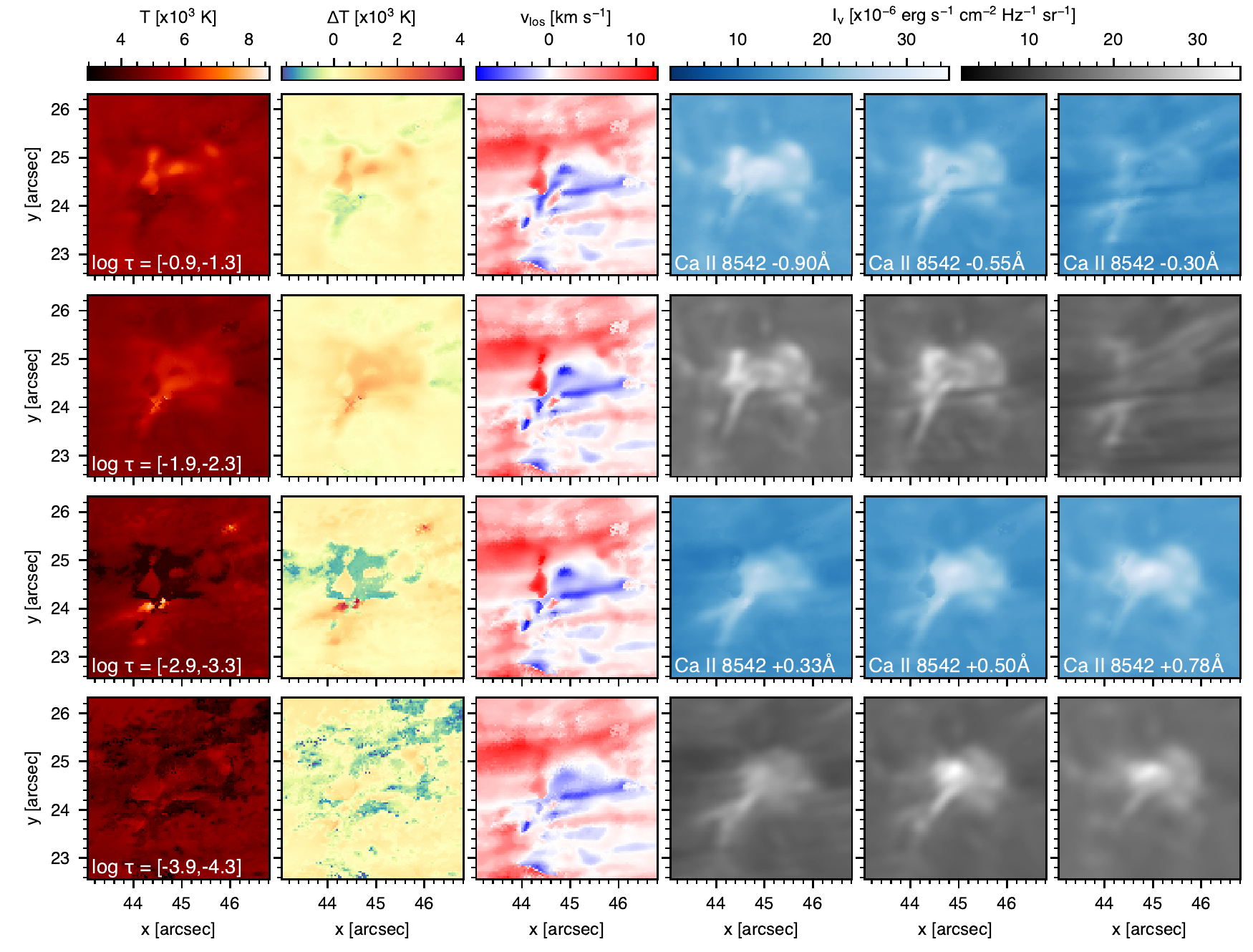}}
  \vspace{-2ex}
  \caption[]{\label{fig:eventA_ca8_maps} %
    Inversion maps of the second snapshot of Event A
    (cf.~Fig.~\ref{fig:subfov}, second row) from the \CaIR-only run at
    several heights in the model atmosphere ({\it three left-hand 
    columns\/}), as well as synthetic and observed intensity images for
    comparison ({\it three right-hand columns\/}).
    {\it From left to right\/}: temperature, the temperature difference with
    respect to the initial input model, line-of-sight velocity, synthetic
    ({\it first and third rows\/}) and
    observed ({\it second and fourth rows\/}) intensity images in the
    wings of \CaIR, at the wavelength offsets indicated in the synthetic image
    panels.
    The \ltau\ height for each row is indicated in the lower left of the
    left-most panels.
    Panels are scaled by column for the three left-hand columns, while all
    panels in the three right-hand columns are scaled to the same values, \ie\
    similar colours in different panels means similar values at different
    heights or wavelength offsets. The intensity panels have been
    multiplied by 1.25 to offset absolute intensity differences with \CaIIK,
    allowing direct comparison with the \CaIR\ panels in
    Figs.~\ref{fig:eventA_maps} and \ref{fig:eventA_mg_maps}.
  }
\end{figure*}

\subsection{Inversions of CRISP and CHROMIS data}
\label{sec:invAB_sst}
\begin{figure*}[bht]
  \centerline{\includegraphics[width=\textwidth]{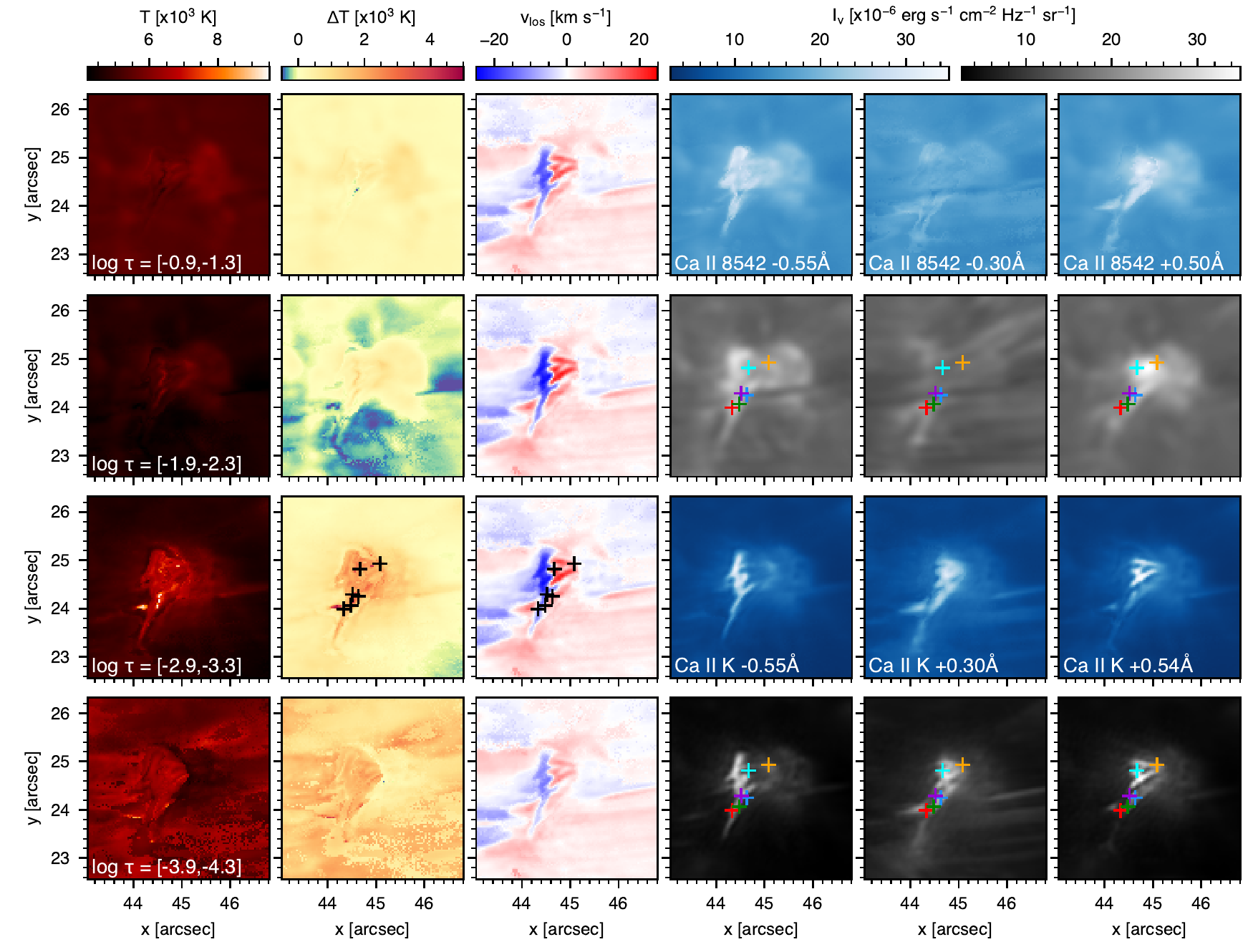}}
  \vspace{-2ex}
  \caption[]{\label{fig:eventA_maps} %
    Inversion maps of Event A considering both \CaIR\ and \CaIIK.
    Format as for Fig.~\ref{fig:eventA_ca8_maps}, except that the right-hand panels in
    the lower two rows now show the synthetic and observed \CaIIK\ intensity
    images (at the specified wavelength offsets) and those in the upper two rows
    (\ie\ \CaIR) have been multiplied by a factor 1.25 to offset the intrinsic
    intensity difference between both calcium lines.
    The coloured plus markers in the right-hand column indicate the locations
    for which similarly-coloured profiles are shown in
    Fig.~\ref{fig:eventAB_profs}.
    For reference, the same markers are overplotted on the third temperature
    difference and line-of-sight velocity maps (\ie\ around $\ltau\tis-3.1$),
    albeit in black for better visibility.
  }
\end{figure*}

We first consider inversions of the CRISP and CHROMIS data of both Events A and
B.
Figures~\ref{fig:eventA_ca8_maps} and \ref{fig:eventA_maps} show inversion maps of
the second snapshot of Event A at four heights in the atmosphere
($\ltau\simeq[-1.1,-2.1,-3.1,-4.1]$, averaged over three nodes around each)
from, respectively, the \CaIR-only inversion and the inversion including both
the \CaIR\ and \CaIIK\ lines.  
Figure~\ref{fig:eventB_maps} shows similar maps of the combined \CaII\
inversions for both Event B snapshots of Fig.~\ref{fig:subfov}.

\paragraph{Event A: \CaIR-only results.}
The right-hand columns in Fig.~\ref{fig:eventA_ca8_maps} show that the general
intensity patterns are well-recovered by the inversion code, including the range
of intensity values (the difference in dynamic range between each set of
synthetic and observed panels is negligible), although the darker patch to
the left of the bright event in the fourth and fifth panels of the
third row clearly shows the code has trouble in some parts of the sub-FOV.
This feature corresponds morphologically well with the bright event top in
the $-$0.90\,\AA\ panels (first and second panels in the fourth column), but
shows unexpectedly a redshift suggesting the code has difficulties
there---likely because of the contribution from the dark fibrils visible in the
$-$0.30\,\AA\ panels (first and second panels in the last column).

The general shape of the \EB\ is also recovered in the temperature maps at
$\ltau\tis-$2, albeit with less of the spatial structure than is visible in the
intensity panels. 
At higher heights the event is practically lost, but at $\ltau\tis-$1 the compact
brightening in the top intensity panels is recovered as an enhanced temperature
of some \dT=1000--2000\,K (at $(x,y)\simeq(44\farcs{2},24\farcs{5})$), with a
hint of enhanced temperature in a semi-circular arc to the left of the main
brightening.
Overall the total temperature reaches order 7500--8500\,K, corresponding to some
\dT=3000--4500\,K over the local ambient temperature.
The highest temperatures rise of roughly \dT=4000\,K is found close to
$\ltau\tis-$3 and its location in the observed plane corresponds to the stronger
brightening at $(x,y)\simeq(44\farcs{3},24\farcs{0})$ in the extended jet
that is visible both in all intensity panels.
The cooler temperatures that cross the event at $\ltau\simeq-$3 and the noisy
temperature maps at $\ltau\tis-$4 are most likely due to the dark canopy fibrils that
are evident in the blue-wing images (in particular those near $-$0.3\,\AA), but
are ill-recovered due to the reduced temperature sensitivity of \CaIR\ at
those heights.

The line-of-sight velocity maps are even more confused.
Disregarding the earlier mentioned artefact, there is still a mix of up- and
downflows, both at the base of the event (at
$(x,y)\simeq(45\farcs{0},24\farcs{5})$) and what would correspond to the
jet-like extension towards the lower left.
The latter is distinguishable to some extent as a blue-shift protrusion flanked
by a small red-shifted feature to its right up to $\ltau\simeq-$3, at
$(x,y)\simeq(44\farcs{5},24\farcs{0})$, which also coincides spatially with
the location of the largest temperature enhancement.

\paragraph{Event A: \CaIR\ and \CaIIK\  results.}
Comparison with Fig.~\ref{fig:eventA_maps} evidences the advantage of
considering multiple diagnostics simultaneously.
In particular the panels at $\ltau\simeq-$3 and $-$4 show much better
defined structures than with \CaIR\ alone. 
In part this is because of the high-resolution CHROMIS \CaIIK\ data displaying
more fine structure than the CRISP \CaIR\ images, however, the model is also
better constrained by including lines formed at somewhat different heights.
The added continuum point from the \CaIIK\ observations further constrains
the temperature at the lowest heights, resulting in a better fit to the lines
and consequently a better constraint of the line-of-sight velocity gradients.
Again, the synthetic images in the first and third rows coincide well with the observed
ones in the second and third, both in terms of feature shapes and dynamic range, suggesting
that also here most profiles are well-fitted.

\begin{figure*}[bht]
  \centerline{\includegraphics[width=\textwidth]{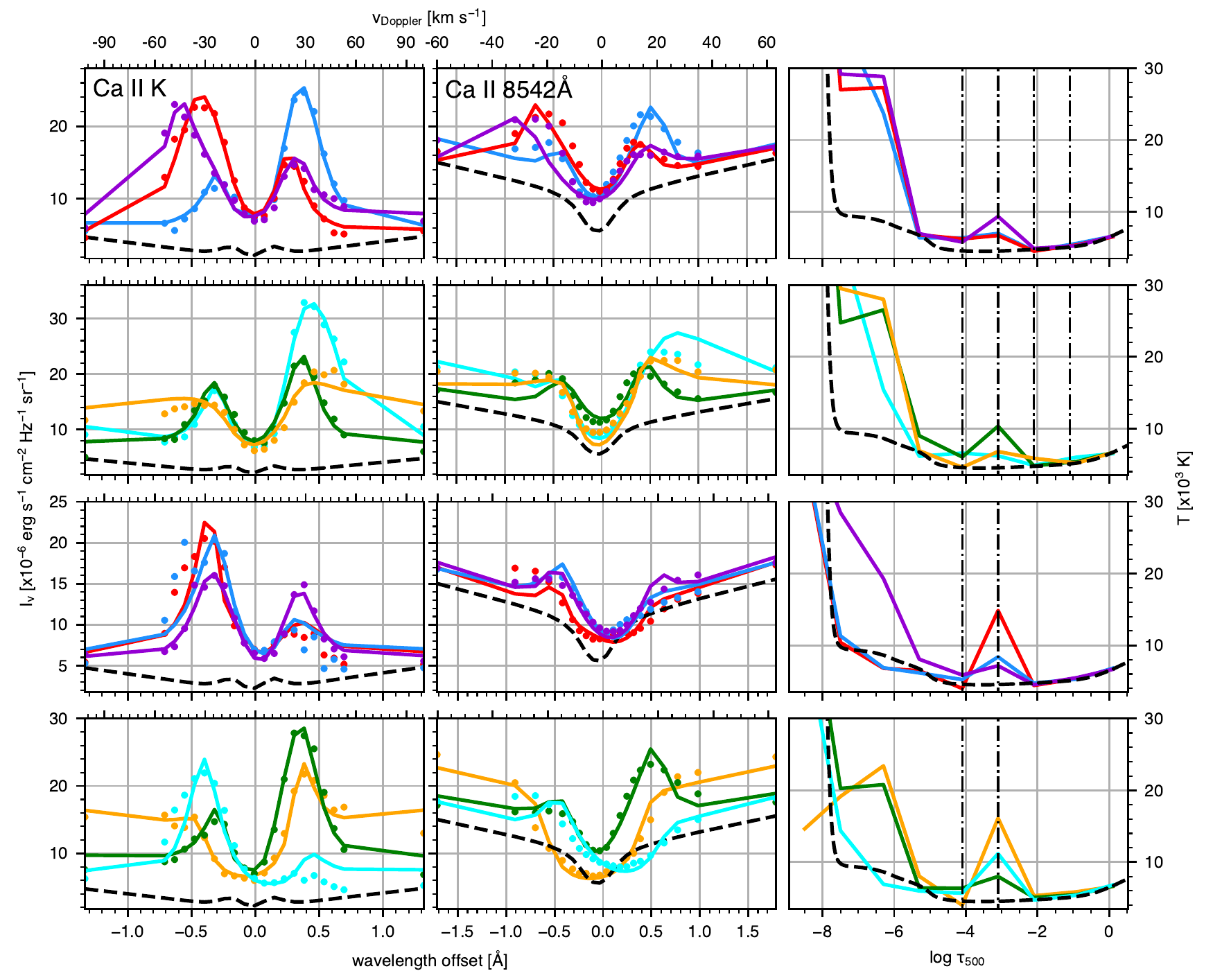}}
  \vspace{-2ex}
  \caption[]{\label{fig:eventAB_profs} %
    \CaIIK, \CaIR\ and temperature profiles for six selected pixels in Events A
    ({\it top two rows\/}) and B ({\it bottom two rows\/}) at the identically coloured
    locations marked in Figs.~\ref{fig:eventA_maps} and \ref{fig:eventB_maps},
    respectively.
    The left-hand and middle panels show fits ({\it solid lines\/}) to the observed profiles
    ({\it filled circles\/}) of the \CaIIK\ ({\it left panel\/}) and \CaIR\ ({\it middle
    panel\/}) lines, respectively.
    The corresponding temperature stratification ({\it right-hand panels\/}) is shown
    using the same colour coding.
    In the latter, the black dashed curve represents the input temperature
    stratification (initialised from FAL-C, for which the \CaII\ panels 
    show the profiles with similar line style for reference), 
    while the vertical dash-dotted lines
    indicate the \ltau\ heights at which the maps in Figs.~\ref{fig:eventA_maps}
    and \ref{fig:eventB_maps} are shown.
  }
\end{figure*}

\begin{figure*}[h]
  \centerline{\includegraphics[width=\textwidth]{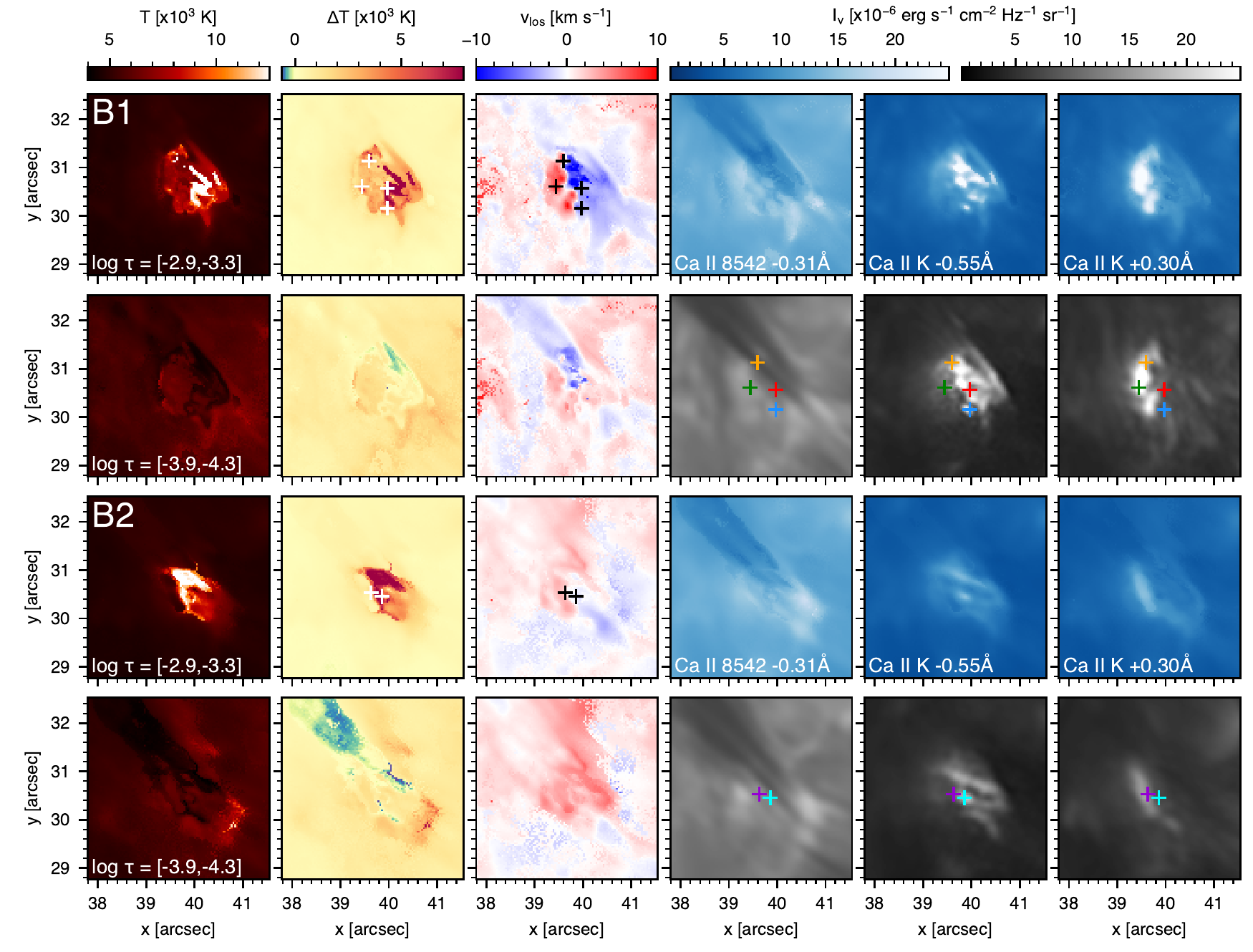}}
  \vspace{-2ex}
  \caption[]{\label{fig:eventB_maps} %
    Inversion maps of Event B based on \CaIR, \CaIIK\ and \FeI\ data, in format
    similar to Fig.~\ref{fig:eventA_maps}, but only showing maps at two
    heights for each Event B1 and B2.
    The top two rows (B1) show maps for the first snapshot of Event B in
    Fig.~\ref{fig:subfov}; the bottom rows (B2) show those for the second
    snapshot. The coloured plus signs mark locations for which \CaII\ profiles are
    shown in the bottom two rows of Fig.~\ref{fig:eventAB_profs}. For the
    orange location in B1 Fig.~\ref{fig:eventAB_mg_profs} also shows
    \MgII\ profile fits.
  }
\end{figure*}

The top two rows of Fig.~\ref{fig:eventAB_profs} highlight this by comparing
single-pixel fits to observed profiles for a number of sampling locations in
Event A (indicated with identically coloured plus markes in Fig.~\ref{fig:eventA_maps}).
While the fitted profiles (solid lines) are not perfect everywhere, generally
good agreement is obtained for both lines, albeit more so for \CaIIK\ than for
\CaIR---likely an effect of assigning more weight to the former.
Of those shown, the orange and cyan profiles have the largest mismatch issues in
one (or both) of the lines, which is likely related to wing asymmetries. 
For instance, the cyan \CaIIK\ profile exhibits a strong blue-over-red
asymmetry, which presumably drives the solution to have a similar asymmetry in the
8542\,\AA\ line; for the orange profile the absence of such asymmetry in the
8542\,\AA\ has lead to underestimate/overestimate the \CaIIK\ red/blue wing and
\Ktwo\ peaks.
Another issue may be the time difference (in this case of 8.8\,s) between the
\CaIIK\ and the \CaIR\ scans, \ie\ the observed profiles are not perfectly
co-temporal even though STiC assumes they are.

While in the \CaIR-only inversion Event A is most clearly visible in the
temperature map at $\ltau\tis-$2, it is not as clear at that height when combining both
calcium lines.
At this height, only a narrow zig-zag shaped temperature enhancement that coincides
spatially with a brightening of similar morphology in the \CaIIK\ blue wing at
$-$0.55\,\AA\ (third and fourth panels in the fourth column) is evident.
Interestingly, this enhanced brightness and temperature coincides with the
boundary of the red-shift signature to the right and blue-shifts to the left.
On the other hand, the strongest temperature enhancements in the combined \CaII\
inversion are reached in similar locations as in the \CaIR-only run. 
The green and purple crosses mark two of such locations and the corresponding
temperature stratifications indicate that some 9500--10,000\,K (up to \dT=5500\,K
over the ambient input temperature) is reached around $\ltau\tis-$3.
In addition, for most samplings a sharp temperature increase is found around
$\ltau\simeq-$5.5, which for some (\eg\ the blue and cyan profiles) represents a
considerable lowering of the transition region, while for others (the red,
purple, orange and green profiles) it seems to be the increase to a
``chromospheric'' temperature plateau at some 2.5--3.0\,\dakK. 

Including \CaIIK\ has also considerably altered the line-of-sight
velocity maps.
The extended jet visible in the intensity images is now clearly recovered as an
elongated blue-shifted feature of some 15--20\,\kms\ towards the observer, while
strong red-shifts of similar magnitude are found at the base of the event.
Such bi-directional jet signature was already implied in the composite \CaIIK\
image of Fig.~\ref{fig:subfov} (second row, third panel), but is now also
confirmed from the inversions.

\paragraph{Event B: \CaII\ and \FeI\ results.}
Figure~\ref{fig:eventB_maps} shows the inversion maps for both snapshots of
Event B, from including both \CaII\ lines and \FeI\,6301.5\,\AA\ (latter not
shown), in similar format as Fig.~\ref{fig:eventA_maps}. 
The upper two rows correspond to the first snapshot (marked B1) and the lower
two rows to the second one (B2).
Again STiC is able to recover the fine structure in the intensity images (but
also the \vlos\ maps), doing slightly better for snapshot B1 than B2, though
with very similar results for both \CaII\ lines.
Overall, for both snapshots the discrepancies are mostly in the \CaIR\ panels: 
in case B1 an imprint of the \CaIIK\ brightenings is visible that is
not there in the observations, while for both
B1 and B2 the dark fibrilar structure overlying the event is relatively
well-reproduced in intensity at the wavelengths shown.
The blob-like substructure is evident for the first snapshot (B1), in particular in
the \CaIIK\ intensity panels (two right-most panels in the first and second
rows).  
The difference in dynamic range is 1--2$\times10^{-6}$\,\cgsint\ at most for all
panels shown, except the red-wing \CaIIK\ images of frame B2 (third panel
in the fifth column), where
the maximum synthetic intensity falls some 4$\times10^{-6}$\,\cgsint\ below the
observed values.
This assessment is supported by the profile fits shown in the lower two rows of
Fig.~\ref{fig:eventAB_profs}, highlighting four pixels from B1 and two from B2.
In some cases the large \KtwoR\ peak appears to drive a stronger decrease in the
\CaIR\ red wing (\eg\ the green profiles) or both lines are overestimated in the
wing emission (\eg\ red and blue, in particular for \CaIIK); in others (\eg\
purple and orange) both lines are well-fitted simultaneously.

The velocity maps for both frames B1 and B2 do not show as clear a
spatially resolved bi-directional jet structure as for Event A.
Rather, the majority of blobs in B1 show either a clear red-shift/blue-shift
of order 10--12\,\kms\ away/towards the observer.
The line-of-sight velocities are strongest around the height where the
temperature peaks, $\ltau\tis-$3.
The dark fibrils seen in the blue wing of \CaIR\ are well-recovered in the
synthetic intensity images and the $\ltau\tis-$4 panel for Event B1 (third
panel in the second row) displays a moderate blue-shift with similar morphology
at that same location.

In terms of temperatures, the inversions suggest even higher values are reached
in Event B than for Event A.
Both snapshots B1 and B2 show hot patches of up to 15,000\,K total
temperature, peaking between
$\ltau\tis-2$ and $-$4 (cf.~Fig.~\ref{fig:eventAB_profs}), and while one would be
hard-pressed to recognise the blob-like morphology from these temperature maps,
the largest temperature enhancements are found at the locations where the
brightest blobs are visible in the \CaIIK\ images.
The temperature stratification shown in the last panel of the lower two rows of
Fig.~\ref{fig:eventAB_profs} is similar to that for Event A: a temperature
increase around $\ltau\tis-$3 and a rise to the transition region or enhanced
chromospheric plateau close to $\ltau\tis-$5, even though for half of the samplings
the transition region rise lies close to that of the input model.
The apparent decrease above $\ltau\tis-$7 for the orange sampling is an effect of
extrapolation beyond the last node with the slope at the last node.

\begin{figure*}[bht]
  \centerline{\includegraphics[width=\textwidth]{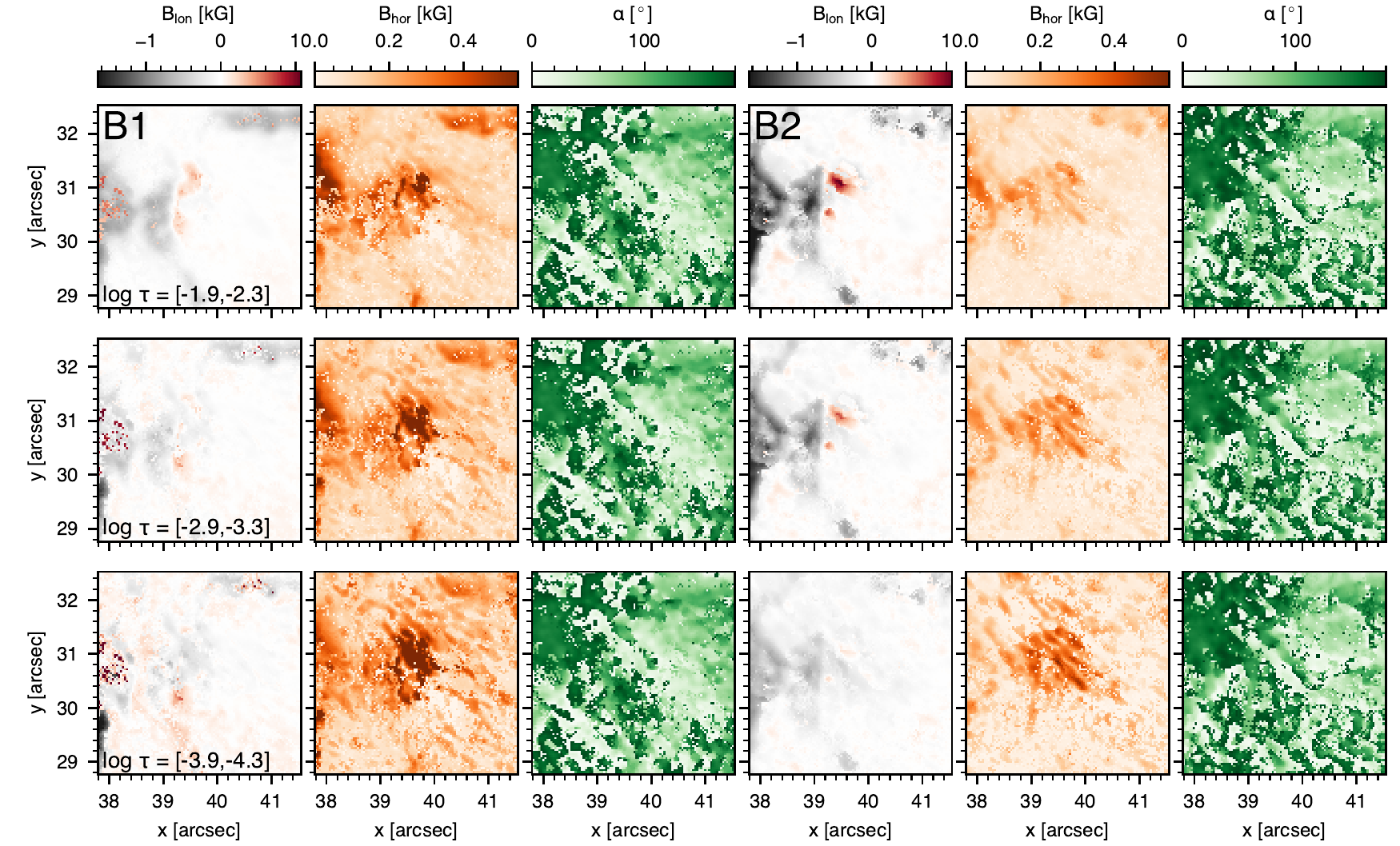}}
  \vspace{-2ex}
  \caption[]{\label{fig:eventB_Bmaps} %
    Inversion maps of magnetic field quantities for both Event B snapshots.
    {\it From left to right\/}: line-of-sight magnetic field strength (with
      positive/negative (\ie\ red/black) corresponding to field oriented
    towards/away from the observer),
    horizontal magnetic field strength and azimuth for Event B1, followed by
    similar maps for B2.
    As only one node was used in azimuth, the corresponding maps
    are identical for the three heights shown.
    The panels have been scaled by column to the same values for both B1 and B2
    to facilitate comparison of the time evolution between them.
  }
\end{figure*}

\subsection{Magnetic fields} 
As for Event A only \CaIR\ polarimetry was available, we allowed fewer degrees
of freedom and the inverted field was consequently not well-defined with height.
However, with the availability of \FeI\ for Event B we added a third node in
line-of-sight magnetic field strength and obtained more reasonable results.
Figure~\ref{fig:eventB_Bmaps} shows the magnetic field inversion maps at the
same \ltau\ heights as Fig.~\ref{fig:eventB_maps} shows the other inversion
parameters, as well as around $\ltau\tis-$1.
While obviously noisy, clear signal is obtained for the longitudinal and
horizontal magnetic fields. 
As only one node was used to fit the azimuth, the maps look the same at all
shown heights.

Especially at the lowest heights the photospheric magnetic field pattern visible
in \FeI\ Stokes $V/I$ (right-hand panels in the lower two rows of
Fig.~\ref{fig:subfov}) is also clearly recovered in the \Blon\ panels.
Going to higher heights the signal gets weaker and more homogenous, as one would
expect from the expansion of the field.
The horizontal magnetic field strength maps are largely devoid of signal and
generally noisy, yet do show enhanced signal at the location where the
brightenings are visible at $\ltau\tis-$3 and $-$4.
For B1 this enhanced horizontal field interestingly overlaps with the location
where the \vlos\ signature flips sign in the top row of
Fig.~\ref{fig:eventB_maps}.
The azimuth maps are noisy at best and do not show a clearly defined structure
coinciding with the event brightening, temperature or line-of-sight velocity
structures, although the values are typically low (below about 40\deg) and
spatially smoother at the locations of enhanced \Bhor.
Considering the changes with time going from B1 to B2 we observe an increase of
the longitudinal magnetic field, mostly at the lower heights \ie\ $\ltau\tis-$2 and
$-$3, while the horizontal fields at the location of the event decrease by
nearly 500\,G (the maximum \Bhor\ values are slightly over 1.1\,kG for B1).
In addition, the B2 maps for \Blon\ at $\ltau\tis-$2 (displaying clear opposite
polarity footpoints) and \Bhor\ at $\ltau\tis-$4 (with enhanced horizontal fields)
are consistent with with a $\cap$-configuration or possibly the shared
horizontal part of a post-reconnection $\cap$- below $\cup$-configuration.

\begin{figure*}[h]
  \centerline{\includegraphics[width=\textwidth]{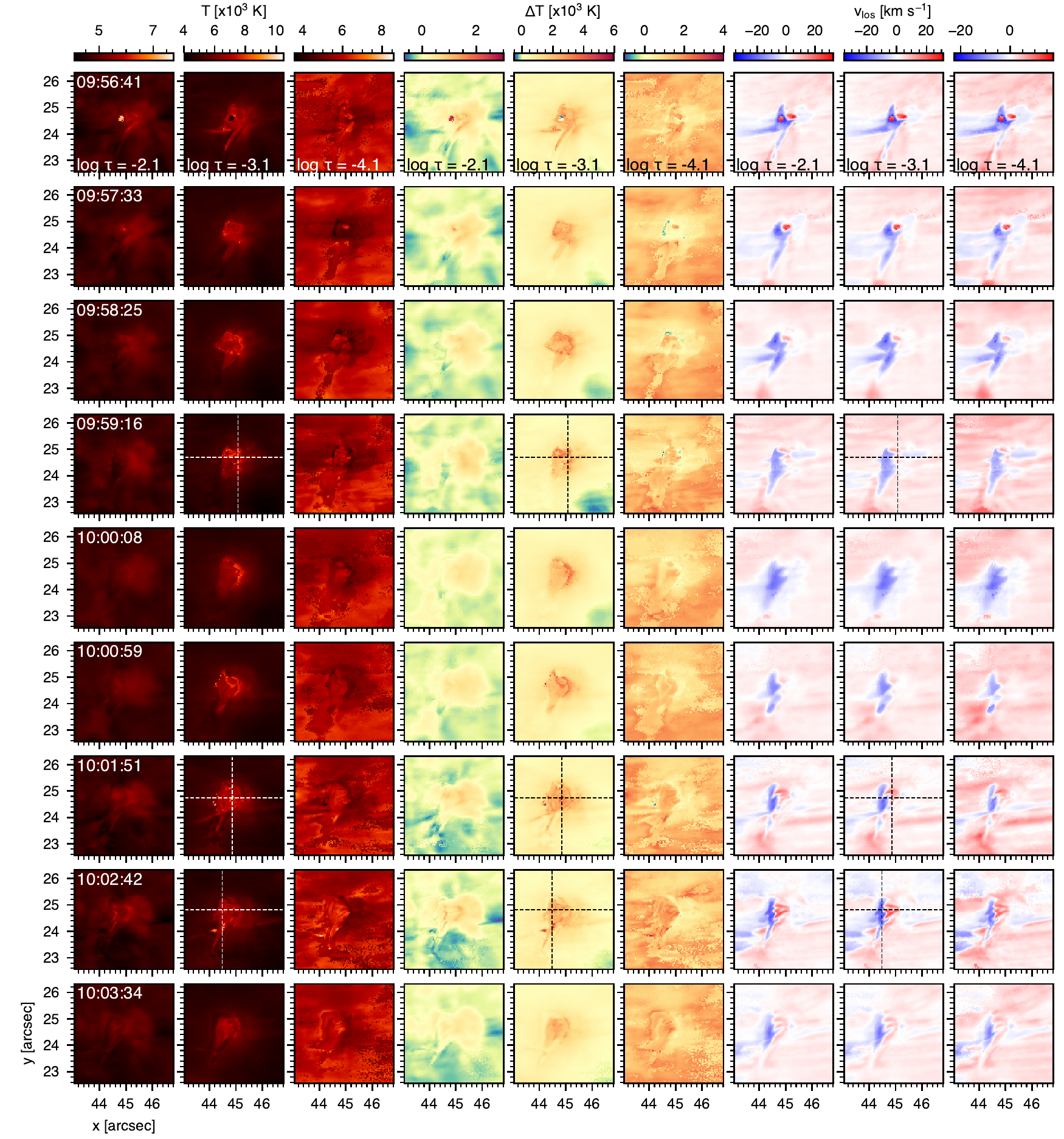}}
  \vspace{-2ex}
  \caption[]{\label{fig:eventA_tmaps} %
    Time sequence of inversion maps of Event A from combining \CaIR\ and
    \CaIIK. From left to right the columns
    show temperature ({\it left-most three columns\/}), the temperature difference
    with the input model ({\it middle three columns\/}) and line-of-sight velocity ({\it right-most
    three columns\/}) at the \ltau\ heights specified in the first row panels.
    The time in UT is indicated in the top left of the first panel of each row.
    The panels have been bytescaled by column (ranges indicated by the colour
    bars at the top of each column).
    The dashed lines in rows 5, 8 and 9 indicate the lines along which
    Fig.~\ref{fig:eventA_tcrossmaps} shows cross-cut inversion maps.
  }
\end{figure*}

\subsection{Time evolution} \label{sec:time}
Let us now consider the time evolution of Event A.
Figure~\ref{fig:eventA_tmaps} shows the top-down inversion maps based on
\CaIR\ and \CaIIK for temperature,
temperature difference and line-of-sight velocity for this event as function of
time at roughly 50\,s intervals.
Throughout this sequence the event stands out in the temperature maps as an
enhancement of a few thousand kelvin up to a total temperature of
8000--10,000\,K, in particular around $\ltau\tis-$3.
At $\ltau\tis-$2 and $-$3 it is clearly hotter than its surroundings, while at
$\ltau\tis-$4 the whole sub-field-of-view appears enhanced in temperature by
\dT=2000--2500\,K and the event starts to blend in.

The bi-directional jet signature so clearly visible in the middle column of
Fig.~\ref{fig:eventA_maps} (and second-to-last row of this figure, at
10:02:42\,UT) can be
observed at several stages during the time evolution, although a line-of-sight
blue-shift of up to $-$25\,\kms\ appears to be the persistent velocity
signature.
In addition, both the top few rows (09:56:41--09:58:25\,UT) and bottom few ones
(10:01:51--10:03:34\,UT) appear to show the blue-shift velocity imprint from
overlying canopy fibrils to the left of the event.
Similar to some of its preceding frames (not shown here), the top row shows
red-shift artefacts embedded in otherwise smooth blue-shifts in the \vlos\ maps
or enhanced and decreased temperatures at $\ltau\tis-$2 and $-$3, respectively,
which corresponds to locations where the maps suffer from worse line fits.

\begin{figure*}[bht]
  \centerline{\includegraphics[width=\textwidth]{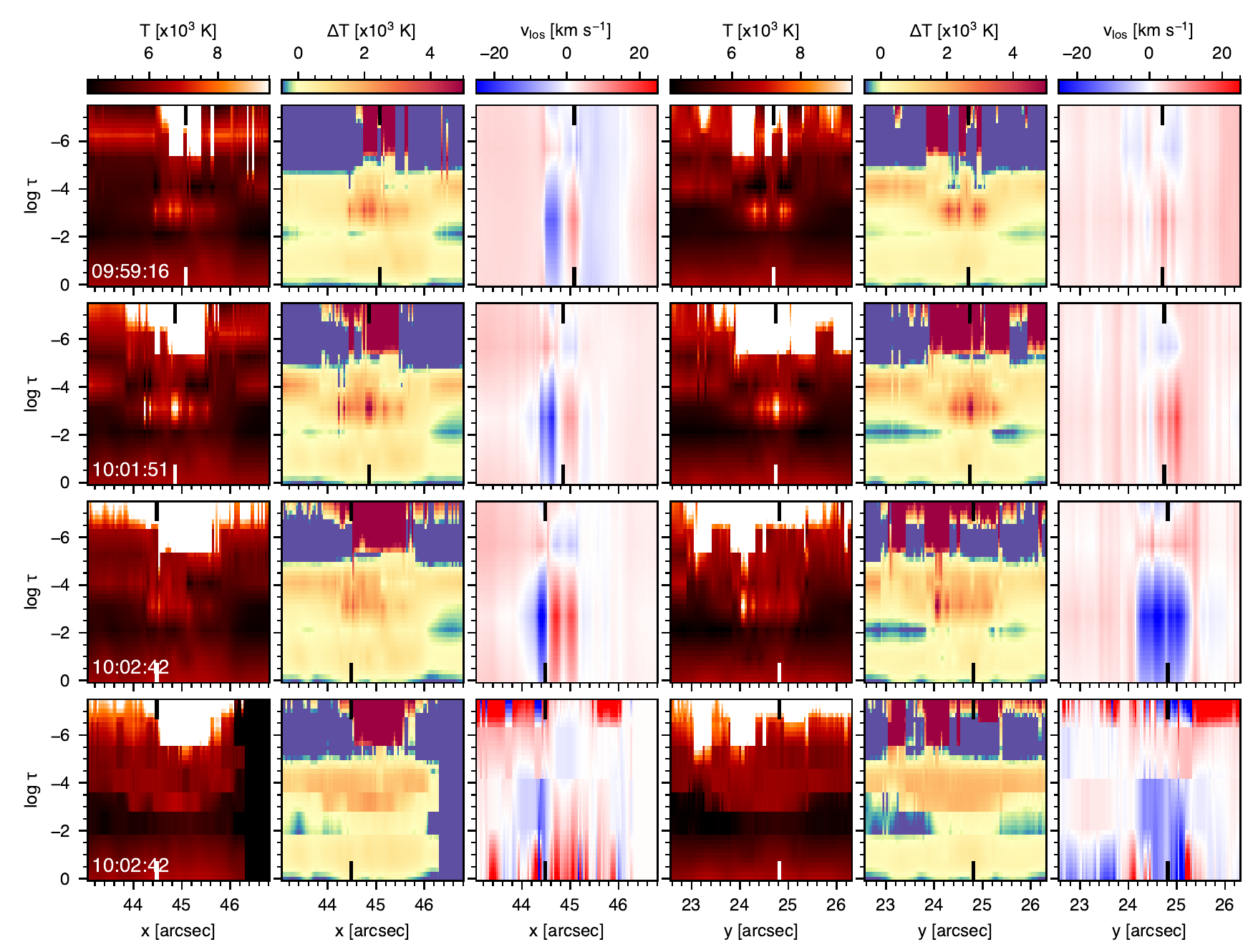}}
  \vspace{-2ex}
  \caption[]{\label{fig:eventA_tcrossmaps} %
    Cross-cut inversion maps of Event A from selected time steps
    highlighted in Fig.~\ref{fig:eventA_tmaps}.
    {\it From left to right\/}: $x$-\ltau\ cut in temperature, temperature
    difference and line-of-sight velocity, followed by the $y$-\ltau\ cuts in the
    same quantities. The panels have been scaled by column, where the
    quantities have been clipped to the respective colour bar values.
    The vertical markers in the $x$-\ltau\ ($y$-\ltau) panels indicate the
    location where they are intersected by the $y$-\ltau\ ($x$-\ltau) cut.
    The lower row displays the cross-cuts of the same frame as in
    the third row, but now from the inversions including \MgII.
  }
\end{figure*}

Figure~\ref{fig:eventA_tcrossmaps} shows cross-sectional inversion maps
for three selected frames from the time sequence in Fig.~\ref{fig:eventA_tmaps}
along the lines indicated in the respective panels; the third and
fourth rows of
maps correspond to the second Event A snapshot of Fig.~\ref{fig:subfov},
where the fourth row is from the inversions including \MgII.
As was already implied by the temperature maps in Fig.~\ref{fig:eventA_tmaps}
and the temperature profiles in Fig.~\ref{fig:eventAB_profs}, the temperature
enhancements related to the \EB/\UVB\ are generally located between $\ltau\tis-$2
and $-$4, peaking around $\ltau\tis-$3.
As indicated before, between $\ltau\tis-$5 and $-$6 we find a sharp temperature
increase to an enhanced chromospheric plateau or due to the actual transition
region overlying the event coming down compared to the FAL-C input (where it lies
close to $\ltau\tis-$8).

The line-of-sight velocity cross-cuts suggest bi-directional flows with the
expected red-shifts below blue-shifts within some of the pixels (cf.~\eg\ the
$x$-\ltau\ cuts for all three frames or the $y$-\ltau\ for the top two).
Interpreting these as the bi-directional reconnection jet within the pixel (\ie\
within the $x$-\ltau\ or $y$-\ltau\ column) would, however, imply the
reconnection takes place above the main temperature increase, as the \vlos\
divergence point is located around $\ltau\tis-$5 in these examples.
Given that the bi-directional signature is clearly separated spatially in the
top-down view, one could imagine the blue-shift should be observed in adjacent
pixels rather than in the same (\ie\ to the top left in these $x$-\ltau\ cuts).
This is the case for all three examples shown, but the extension of both
signatures over $\ltau\tis[0,-4]$ is somewhat puzzling, \ie\ even though opposite
signatures in adjacent pixels is expected, we would also expect the blue-shifts
to be found predominantly at comparatively higher heights than the red-shifts.
To some extent this is the case in the $x$-\ltau\ \vlos\ panel of the
first frame (top row, third panel), where a blue-shift can be found to the top
left (around $\ltau\tis-$4.5) of the stronger red-shift, although this again places
the outflow point above the main heating location.
The same cut for the next two frames shows this much less clearly (if at all); the
blue-shifts left of the red-shifts extend only marginally higher than the
red-shifts.
An issue that could play a role here is the limited number of \vlos\ nodes used
in the inversions, thereby preventing STiC from finding a consistent solution
that places the outflow point lower down.
Increasing the number of nodes may have alleviated this, however, as pointed out
previously increasing the number generally resulted in worse fits to the \CaII\
lines, hence our choice not to do so.
The fourth row, from inversions including \MgII, is further discussed in the
following section, but we note here that the general temperature and velocity
patterns are largely the same between the inversions with and without that
\MgII\ lines.

\begin{figure*}[bht]
  \centerline{\includegraphics[width=\textwidth]{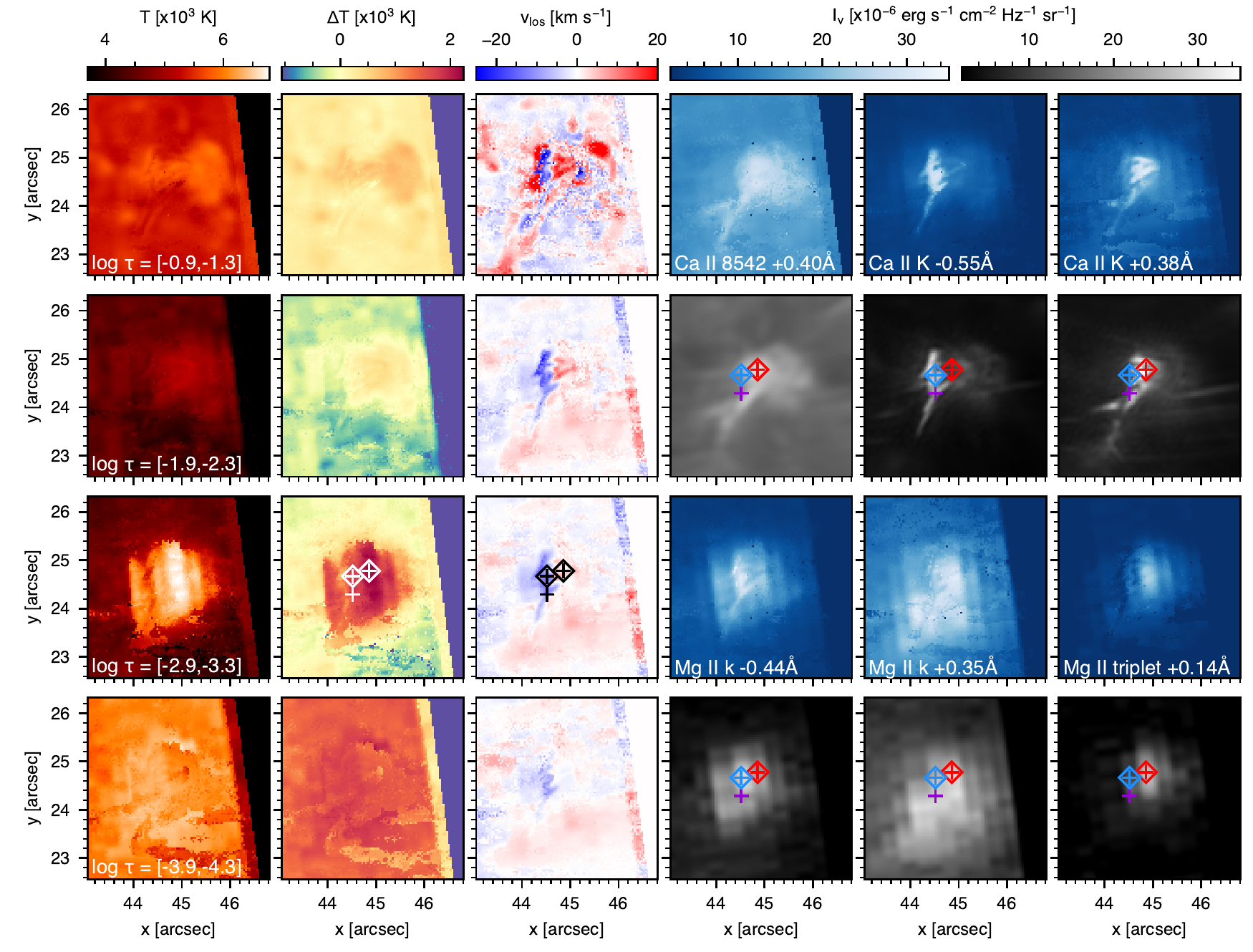}}
  \vspace{-2ex}
  \caption[]{\label{fig:eventA_mg_maps} %
    Inversion maps of Event A considering both \CaII\ lines and \MgIIhk. 
    Format as for Fig.~\ref{fig:eventA_maps}, except that the six panels in
      the lower right now show (from left to right) an image of the
    \MgII\,\ktwoV, \ktwoR\ and the red wing of the subordinate \MgII\ triplet. 
    The \CaIR, \MgIIk\ and \MgII\ triplet images have been multiplied by 1.25,
    7.5 and 10, respectively, to roughly offset the intrinsic intensity
    difference with the \CaIIK\ line and all panels in the three right-hand
    columns have been scaled to the same values subsequently 
    (these are the same values as in Figs.~\ref{fig:eventA_ca8_maps} and
    \ref{fig:eventA_maps}).
    The triangular area to the (upper) right of the inversion maps has values
    set to zero as these fall outside the IRIS raster.
    The coloured plus markers in the second and fourth rows mark the
    locations for which profiles are shown in the top two rows of
    Fig.~\ref{fig:eventAB_mg_profs}; the coloured diamond markers those
    locations for which Fig.~\ref{fig:siiv_invprof} shows profile fits including
    \SiIV.
    The purple cross marks the same location as the purple cross
    in Fig.~\ref{fig:eventA_maps}.
  }
\end{figure*}

\subsection{Inversions of combined SST and IRIS data} 
\label{sec:invAB_sst_iris}
We have seen earlier that including additional diagnostics generally constrains
the model atmospheres better and given that \MgIIhk\ are typically formed
somewhat higher than \CaII\ (cf.~\eg\ Fig.~1 from
\citetads{2013ApJ...772...90L} 
or Fig.~15 from
\citetads{2018A&A...611A..62B}), 
this should help the inversions.
Moreover, finding an atmospheric model that can explain also the UV part of an
\EB\ is of great interest.

When including IRIS the instrumental resolution differences are, however,
non-negligible and a likely source of errors: already between CRISP \CaIR\ 
and CHROMIS \CaIIK\ there is a factor 2 difference in resolution, going up
to over a factor 8 when comparing CHROMIS and these IRIS raster observations.
As we chose not to sacrifice resolution, the IRIS profiles have been
interpolated to the CHROMIS pixel scale, \ie\ a single IRIS spectral profile is
spread out over many SST pixels.
Consequently, STiC considers these one-to-one, while in reality many SST
profiles should be contributing to the atmosphere that explains the single IRIS
profile.
Without a well-defined spatial point spread function for either the SST
or the IRIS spectrograph it is impossible for STiC to take the resolution
difference into proper account. 

Also, as with the previous inversions, additional uncertainty arises from the
acquisition time difference between the different instruments.
On both days the time difference with IRIS can be as large as nearly 11\,s.
While this effect can be minimised, it may still play an important role,
particularly when considering the fast evolution of the \CaIIK\ substructure
where some plasmoid blobs are sometimes only visible for a single frame.

Notwithstanding these issues, the solution to which inversions converge is not
significantly different from the runs without \MgII, as we show below.

\begin{figure*}[h]
  \centerline{\includegraphics[width=\textwidth]{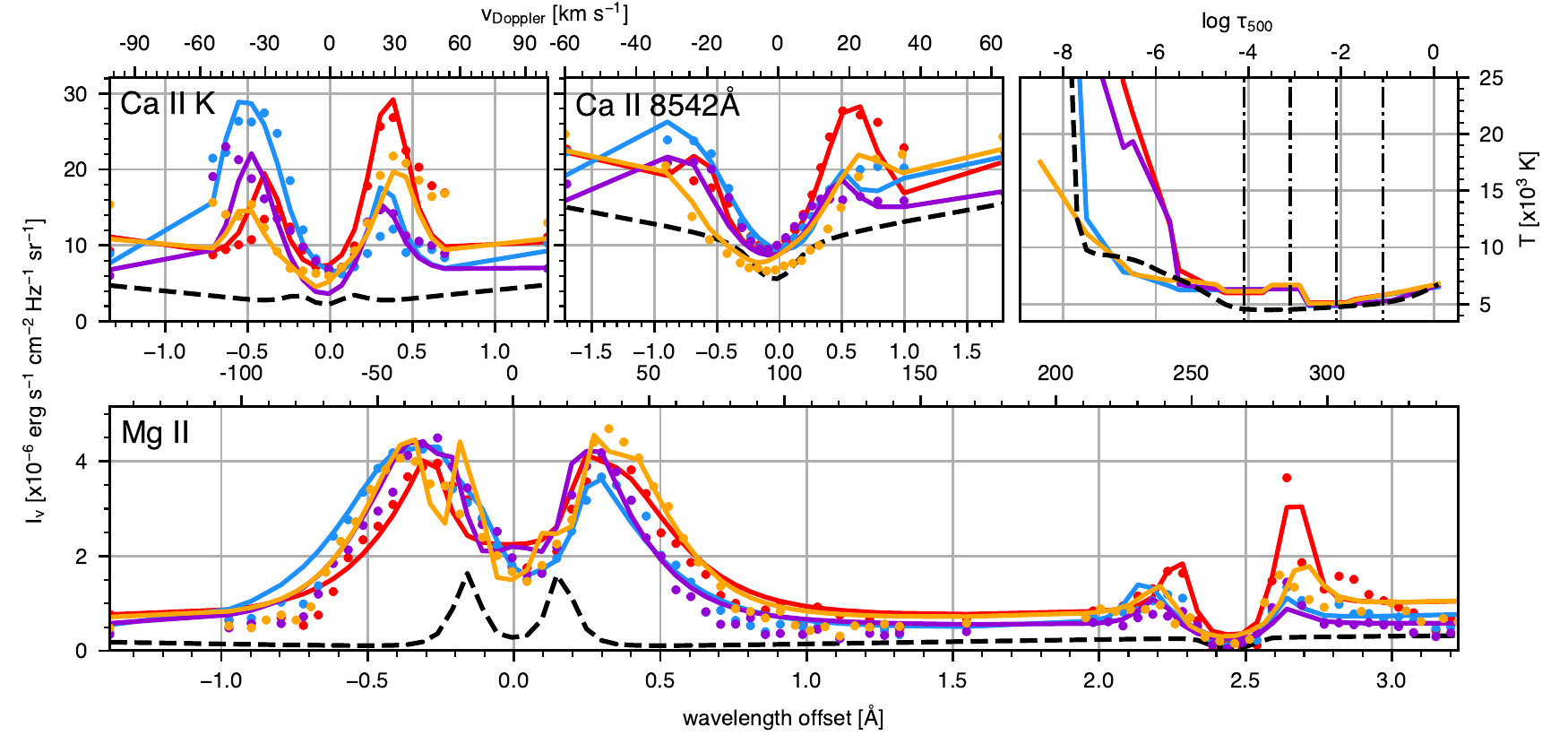}}
  \vspace{-2ex}
  \caption[]{\label{fig:eventAB_mg_profs} %
    \CaIIK, \CaIR, \MgII\ and temperature profiles for selected pixels in event
    A ({\it red, blue and purple\/}) and Event B ({\it
    orange\/})
    at the similarly-coloured locations marked in the right-hand panels of
    Figs.~\ref{fig:eventA_mg_maps} and \ref{fig:eventB_maps}, respectively. 
    Format simlar as for Fig.~\ref{fig:eventAB_profs}, except that the
    second row now shows the \MgIIk\ and \MgII\ triplet lines (the zoom-in has
    been chosen such to allow easier comparison of the observed and inverted
    profiles, omitting \MgIIh\ because it behaves similarly to \MgIIk).
    The input observed \MgII\ profiles are shown with every other point to
    avoid cluttering the plot.
  }
\end{figure*}

\subsubsection{Combining \CaII\ and \MgII} 
\label{sec:invAB_sst_iris_mg}
Figure~\ref{fig:eventA_mg_maps} presents in similar format as before the
inversion results from including \MgII\ along with the \CaII\ lines for Event A.
Comparing Figs.~\ref{fig:eventA_maps} and \ref{fig:eventA_mg_maps} for Event A
we see that at lower heights ($\ltau\tis-$1 and $-$2) the temperature maps look
very similar between the runs without and with \MgII\ (a hint of the fine
structure imprint from \CaIIK\ remains visible in the \MgII\ results).
In fact, at both heights the range of temperatures is similar between the runs
and also the average temperatures fall within 50--250\,K of each other.
Also for both, the line-of-sight velocity maps show clearly the bi-directional
jet in most panels of the middle column, most clearly so at $\ltau\tis-$2.
However, when including \MgII\ the bi-directional pattern is much more noisy at
$\ltau\tis-$1.

The differences are more pronounced higher up.
On one hand, the contribution from the IRIS-observed \MgII\ is evident in the
temperature maps around $\ltau\tis-$3 through a more dispersed temperature
enhancement, devoid of much of the substructure that was visible when
considering only SST data.
At $\ltau\tis-$4 the event nearly blends into the background in the
temperature maps, with the whole sub-FOV displaying a more or less homogeneous
temperature enhancement of roughly \dT=2000\,K over the input temperature.
Also, while in the \CaII\ run the bi-directional velocity signature remains
clearly visible throughout the inverted atmosphere, when including \MgII\ this
signature is more pronounced in the sense that the redshifts are stronger at
lower heights, likely due to the contribution from the \MgII\ triplet (cf.~the
lower right intensity panels), and disappear almost entirely at $\ltau\tis-$4.
This is also reflected in the line-of-sight velocity cross-cut pattern
differences between the third and fourth rows of
Fig.~\ref{fig:eventA_tcrossmaps}, the latter from inversions including \MgII.
For instance, the strong red-shifts in the $x-\ltau$ cut (third panel of the
third row) do not extend as high when including \MgII\ (fourth row) and in the
latter the adjacent blue-shift (at $x\simeq44\farcs{2}$) is also largely
concentrated between $\ltau\tis-$2 and $-$4, rather than extending all the way from
\ltau\tis0 to $-$4.
This fits with the expectation that the blue-shifts should be found at
comparatively higher heights than the red-shifts.

Figure~\ref{fig:eventAB_mg_profs} shows the \CaII\ and \MgII\ fits with
corresponding temperature profiles for identically coloured selected pixels in
Figs.~\ref{fig:eventA_mg_maps} (red, blue and purple; Event A) and
\ref{fig:eventB_maps} (orange; Event B).
The fits are generally good, although getting agreement in these three lines
simultaneously is more challenging than for the two \CaII\ lines alone.
For the examples from Event A, the profile asymmetries are strongest in \CaIIK\
and the brighter \Ktwo\ peak is also typically the one that is better fitted
(cf.~\eg\ the red and blue profiles), while the \Kthree\ core is sometimes too
dark (\eg\ purple and orange profiles), and in some cases the line wings
are not bright enough (orange profile).
By comparison, the fits to \CaIR\ show generally fewer discrepancies with the
observations than \CaIIK (except for the orange sampling).
As before, it is however possible that differences in the magnitude of the
asymmetries between these lines may affect the fitting of one of the line wings.
For \MgIIk\ the \ktwo\ peaks and inner wings are typically well-fitted,
while showing more (though still minor) issues in the \kthree\ core and further
out in the wings, beyond $\pm$0.75\,\AA, the latter being usually too bright
compared to the observations.
The \MgII\ triplet and its asymmetries are also generally well-reproduced.

Considering the temperature stratification, the peak temperatures of order
1--1.5$\times$10$^{4}$\,K that resulted from the \CaII\ inversions are
interestingly not found when including \MgII. 
Rather the temperature enhancement is a moderate \dT=2500\,K over the ambient
temperature at $\ltau\tis-$3 and for both events the stratification shows an
extendend plateau over a range of $\Delta \ltau \simeq\ 2.5$ upward from there.
This is also evident from the temperature cross-cuts
(cf.~the lower row of Fig.~\ref{fig:eventA_tcrossmaps}).
In particular the purple profile for Event A (or the orange one for Event B),
corresponding to the same sampling location for which the purple (orange) profile
in the first (fourth) row of Fig.~\ref{fig:eventAB_profs} is shown, now only
reaches a total temperature of some 6500\,K.
We discuss these differences further in Section~\ref{sec:discussion}, but note
already here that this is likely an effect of the pixel size difference between
the IRIS and SST data.

\subsubsection{Challanges posed by \SiIV} 
\label{sec:invAB_sst_iris_si}
As defined in 
\citetads{2018SSRv..214..120Y}, 
the primary identification of UVBs is through their enhanced and broadened
\SiIV\ lines and properly reproducing this diagnostic is therefore important for
a complete description of these events.
Unfortunately, at this point this is easier said than done and after several
tests we decided to refrain from inverting maps including the \SiIV\ lines.

\begin{figure*}[bht]
  \centerline{\includegraphics[width=\textwidth]{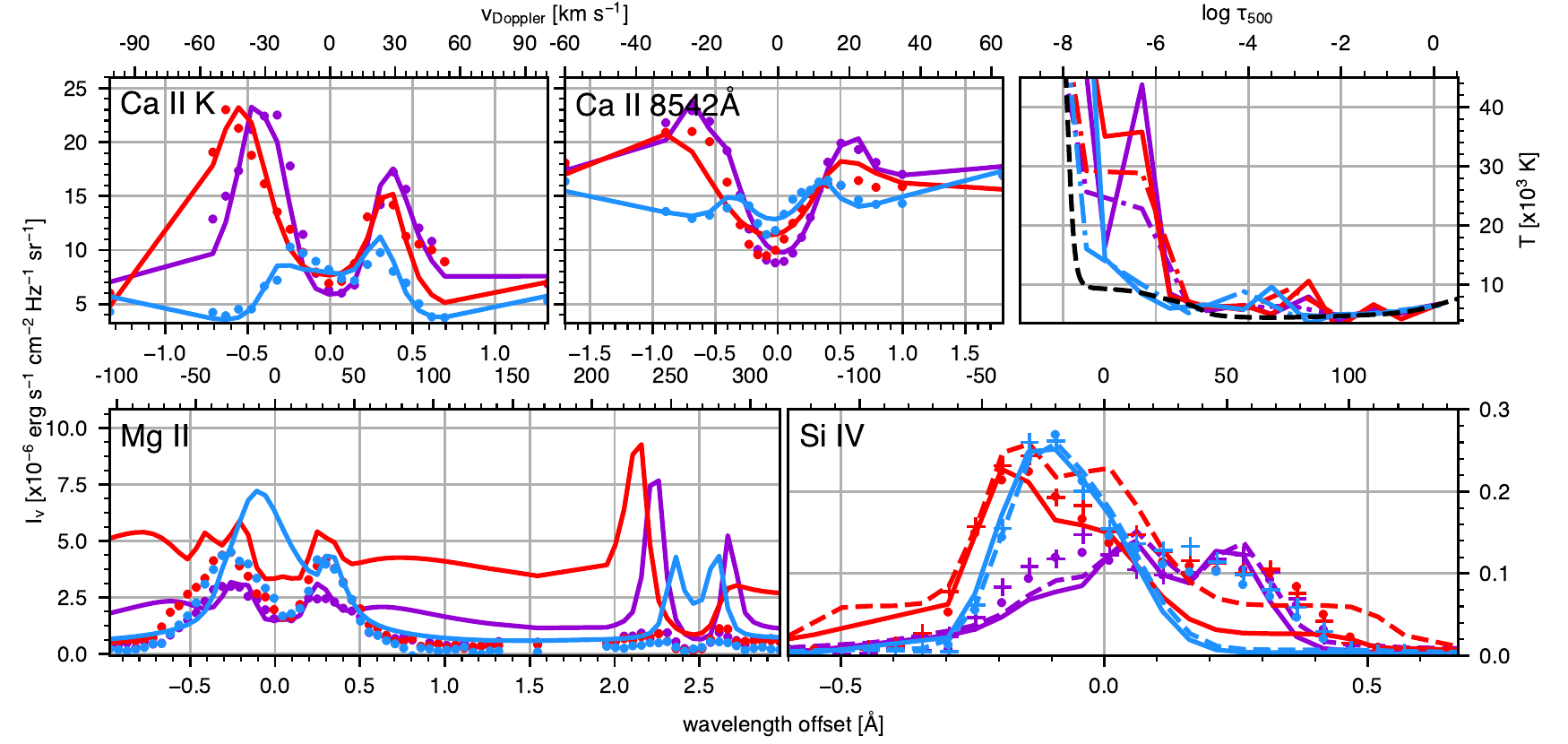}}
  \vspace{-2ex}
  \caption[]{\label{fig:siiv_invprof} %
    Observed and inverted \CaIR, \CaIIK, \MgII, \SiIV\ profiles, as well as the
    corresponding temperature stratification, for selected
    pixels in Event A (highlighted with diamond markers in
    Fig.~\ref{fig:eventA_mg_maps} for the red and blue profiles; the purple
    profiles corresponds a pixel in the middle of the \SiIV\ brightening in the
    earlier Event A snapshot shown in the top row of Fig.~\ref{fig:subfov}).
    Format similar to Fig.~\ref{fig:eventAB_mg_profs}, except that the lower row
    contains an additional panel with the \SiIV\,1394\,\AA\ (1403\,\AA) line,
    where the observed profiles are shown as dots (plus markers) and the fits as
    solid (dashed) lines.
    The observational and synthetic \SiIV\,1403\,\AA\ data have been multiplied
    by 2 to offset the intrinsic intensity difference with 1394\,\AA\ under
    optically thin conditions.
    The temperature profile panel also includes dash-dotted lines for each
    sampling indicating the result from the previous inversion cycle (\ie\ SST data only).
  }
\end{figure*}

Figure~\ref{fig:siiv_invprof} evidences why.
This figure shows the profile fitting results of single pixel inversions at the
locations marked with diamonds in the top rows of Fig.~\ref{fig:subfov}
(two of which are also overplotted in Fig.~\ref{fig:eventA_mg_maps}).
Apart from the \CaII\ and \MgII\ lines as shown before, a fourth spectral panel
now displays the \SiIV\ lines with the 1394\,\AA\ (1403\,\AA) observations and
fits as dots and solid lines (plus markers and dashed lines), respectively.
The \SiIV\,1402\,\AA\ profiles have been multiplied by 2 to account for the offset
between the two \SiIV\ lines when formed under optically thin conditions: comparison of
the dots and plus markers for each sampling line pair indeed suggests they
likely are, as the \SiIV\,1394\,\AA\ to 1403\,\AA\ ratio is close to 2 for
the purple and blue samplings, while the red one is clearly non-thin given a
line ratio of 1.7.

The samplings shown were selected to test fitting of \SiIV\ profiles with
varying degrees of complexity (blue-asymmetry for both blue and red
samplings versus more ragged-top purple sampling profiles).
While STiC may be able to reproduce the general broadening, enhancement and
asymmetry of the \SiIV\ lines, this results in completely off profile fits for
\MgII\ in particular.
Of both \SiIV\ lines, the 1403\,\AA\ line (plus markers and dashed lines) is
sometimes considerably better fitted, \eg\ the red sampling. 
Even though not being perfect, the red-shift side ``plateau'' mimicks the
observations better for 1403\,\AA\ than the solid line follows the 1394\,\AA\ dots.
This further supports the non-thin formation already implied by the line
ratio departure from 2.
For the blue sampling both lines are fitted equally bad, in the sense that the
hump on the red-shift side is not fitted at all, but the fit retains a rather
Gaussian shape while recovering the peak intensity.
In contrast, the purple sampling shows the best fits for both lines of these three
examples, even though it does not fully reproduce the observed intensities
between about $-$50\,\kms\ and the nominal line centre.

The temperature profiles (top right panel, solid coloured lines) do show some
changes with respect to the inversion results when only SST diagnostics were
considered (dashed coloured lines).
The temperature peaks that after cycle 2 were already there close to the input
temperature minimum (\eg\ blue and red) get shifted to lower heights by
about $\ltau\tis0.5$, while also increasing by a few thousand kelvin.
For the purple sampling the temperature enhancement close to the temperature
minimum was less pronounced when considering only SST data, but gets a
noticeable peak now just below $\ltau\tis-$3 when including both \MgII\ and \SiIV.

The behaviour at higher heights is also changed. 
For instance, where the red and purple samplings previously showed an increase to
a chromospheric plateau of about 3\,\dakK\ and 2.5\,\dakK, respectively, the
red temperature profile now shoots up to a plateau over 3.5\,\dakK\ and the
purple profile shows a pronounced peak up to 4.3\,\dakK\ at $\ltau\simeq\,-6.3$.
The blue sampling does not show such enhanced temperature between $\ltau\tis-$7
and $-$6, but the transition region temperature rise starts about
$\ltau\tis0.5$ lower when including both \MgII\ and \SiIV.
It should be noted, however, that the inversions with \SiIV\ were run with 13
nodes in temperature, two more than the second cycle inversions, which may
explain both the slight shift of the low temperature peaks (since the nodes are
slightly shifted), and the ability to resolve the pronounced temperature
peak at $\ltau\simeq\,-6.3$ for the purple sampling (considering that the purple
dashed temperature profile already showed a high temperature chromospheric
plateau).

All in all, while both \CaII\ lines are also well-fitted for all three samplings
(and better so than \SiIV), it is clear that in general agreement cannot be
reached for all diagnostics simultaneously with the proposed node models (\MgII\
being the most difficult to reconcile). 
Considering the temperature profiles this is likely due to the temperature peaks
to some 10$^{4}$\,K between $\ltau\tis-$3 and $-$4, which (as shown in the first
part of this section) were suppressed when considering \CaII\ and \MgII\
together.
Other issues that may play a role include the limitation to pixel-by-pixel
atmospheres (\ie\ restriction to 1.5D), the difference in data resolution,
non-zero acquisition time differences for the spectra and the limited number
and/or placement of temperature inversion nodes.  
We discuss these further in Section~\ref{sec:discussion}, but address the latter
shortly here.

\subsubsection{Effects of sharp temperature enhancements} 
\label{sec:siiv_spikes}
Certain temperatures need to be reached in order to yield any measurable
intensity increase in \SiIV, but because of the node distribution such temperature
bump will typically be broad in \ltau\ corresponding to significant heating over
a large height range.
This is probably one of the reasons for the overestimated enhancement of \MgII\ 
seen in Fig.~\ref{fig:siiv_invprof}. 
Considering the observation of plasmoid-like blobs and indications from
numerical studies (\eg\
\citeads{2018ApJ...852...95N}; 
\citeads{2018arXiv180405631N}) 
that in some configurations the plasmoids may have associated confined slow- and
fast-mode shocks reaching sufficiently high temperatures to explain \SiIV\
emission, it may be that this emission is very localised and hence impossible to
resolve with the current node representation.

\begin{figure*}[bht]
  \centerline{\includegraphics[width=\textwidth]{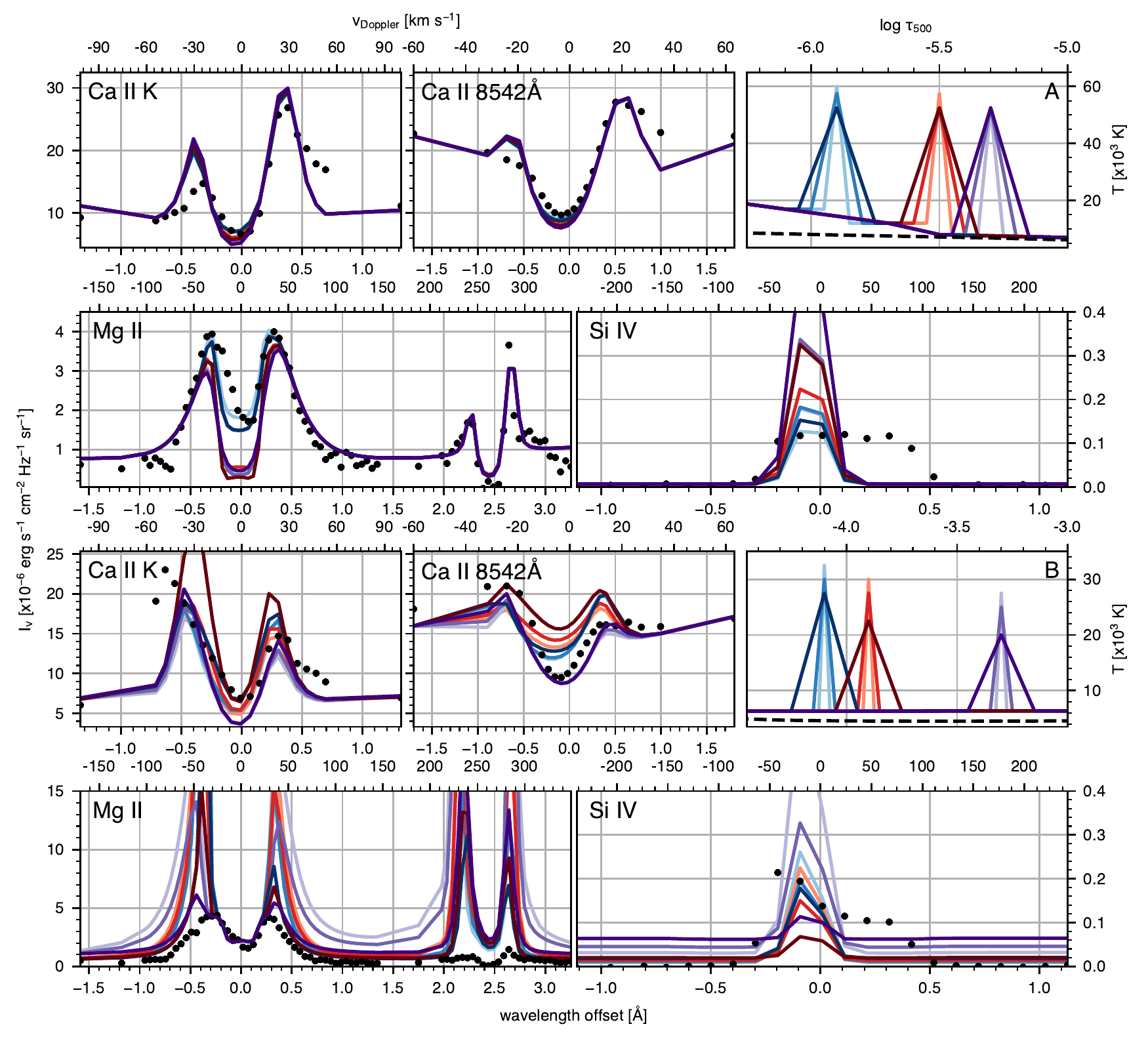}}
  \vspace{-2ex}
  \caption[]{\label{fig:siiv_synthprof} %
    Spectral synthesis of \CaII, \MgII\ and \SiIV\ based on modified inversion
    atmospheres of two well-fitted pixels of Event A.
    Temperature peaks have been added at three \ltau\ heights
    (differentiated by colour) with various peak heights and base widths
    (differentiated by colour shades, increasingly darker for increasing width
    and/or decreasing peak height).
    Subfigure A ({\it top two rows}) shows the results for placing such
    enhancement of \ltau-width 0.1--0.3 ({\it from light to dark shade\/})
    somewhere between $\ltau\tis-$5.9 and $-$4.9. 
    Subfigure B ({\it bottom two rows}) offers a similar display for spikes of
    \ltau-width 0.05, 0.1 and 0.3 ({\it from light to dark\/}) at heights
    between $\ltau\tis-$4.1 and $-$3.1. 
    The underlying observations are from different pixels for the two
    subfigures.
    Each subfigure has similar format as Fig.~\ref{fig:siiv_invprof}, with the
    spectral panels showing the observations as black dotted profiles and
    synthesis output as coloured solid lines.
  }
\end{figure*}

As a first step towards further investigating this possibility, we
forward-modeled the emergent intensity in \SiIV, \MgII\ and \CaII\ assuming a
sharp temperature enhancement at specific heights, sharper than our inversion
node placement would allow to recover.
We tested a grid of temperatures (at $\Delta T=2500\,K$), peak locations (at
$\Delta \ltau\tis0.2$) and base widths ($\ltau\tis0.05$, 0.1, 0.2 and 0.3), but here
only show a sub-selection to highlight certain effects of varying these three
parameters.
Figure~\ref{fig:siiv_synthprof} presents the results of this spectral synthesis for
several examples of single temperature spikes on top of the inverted
temperature profiles based on \CaII\ and \MgII\ data that produce \SiIV\ emission
within an order of magnitude of the observed profile in terms of peak intensity,
where the top (bottom) subfigure shows the effect of localised temperature
enhancements between $\ltau\tis-$5.9 and $-$4.9 ($-$4.1 and $-$3.1).
It is clear that regardless of the input temperature profile, the broadening of
the \SiIV\ is not well-reproduced, yet this is not surprising: for one, the
velocity components (both line-of-sight and non-thermal) were not modified from
the preceding inversion output, but more importantly, in optically thin
formation the width is only influenced by microturbulence and velocity gradients
over the formation region and since the latter is purposefully narrow, the
profile width will be relatively small as well.

Considering the higher altitude perturbations (\ie\ top subfigure) first, adding
the shown temperature spikes does not considerably change the \CaII\ or \MgII\
profiles, except the \kthree\ core of the latter when the spike is located below
about $\ltau\tis-$5.5 (orange-red and purple profiles). 
Comparing identically coloured temperature and \SiIV\ profiles, we see that an
increased base width at the same height and peak temperature (\eg\ full and dark
red, as well as all three purple profiles) causes an increased \SiIV\ peak
intensity.
This is to be expected as the volume that is exposed to the heating is larger,
thus resulting in a stronger \SiIV\ emission. 
At the same time this does not appear to affect its line width, \ie\ the
full-width-at-half-maximum remains essentially unchanged.

The low-altitude perturbations (bottom subfigure) have an effect on all lines,
but most conspicuously on the \CaIR\ core, \MgII\ peaks and wings and the \SiIV\
line.
\CaIIK\ shows similar profiles regardless of the temperature spike location,
height and width, although the \Kthree\ core is darker for the lower location
temperature spikes and the \Ktwo\ peaks are more enhanced for the higher located
spikes.
The higher location spikes (\ie\ close to $\ltau\tis-$4) affect both the \CaIR\
core---which is considerably enhanced compared to the observations---and the
\MgIIk\ and \MgII\ triplet peaks.
On the other hand, the lower location temperature spikes (purple profiles) show
a larger influence on the (quasi-)continua outside the \MgII\ and \SiIV\ lines,
overestimating their intensity, sometimes by more than an order of magnitude.

As one would expect, at higher heights one generally requires higher
temperatures than at lower heights to get similar \SiIV\ emission, but in either
case similar profiles can be obtained from either a narrow but tall or a broader
but lower temperature spike.
The narrowest of temperature spikes can indeed reproduce specific \SiIV\
intensities at (or close to) the nominal line core, but better profile
correspondence likely requires modifying other parameters, such as the
line-of-sight velocity.
In addition, the temperature spikes at lower heights pose problems for the
\MgII\ emission in particular, but depending on the height also for the \CaIR\
core intensity; this may explain why the inversions favoured a solution where
the temperatures were enhanced closer to $\ltau\tis-$6, rather than around the
heights where \CaII\ inversions suggested the temperature increase to be.
Hence, it is clear that this requires further study, yet as this likely requires
a different approach we defer further investigation into the \UVB\ \SiIV\ line
formation to a future study.

\section{Discussion}\label{sec:discussion}
\subsection{Temperature stratification}
\label{sec:discussion_tstrat}
The observation of \SiIV\ and \CII\ emission sometimes correlated and co-spatial
with typical \Halpha-observed \EBs\ has added an additional layer of complexity
requiring explanation and also led to a heated debate over the past few years on the
temperatures these events may reach.
Prior to the IRIS launch, \EB\ temperature estimates ranged anywhere between 
a few hundred to a few thousand kelvin over the ambient temperature at close to
temperature minimum heights, generally based on 1D semi-empirical
modelling
(\citeads{1983SoPh...87..135K}; 
\citeads{2010MmSAI..81..646B}; 
\citeads{2013A&A...557A.102B}; 
\citeads{2014A&A...567A.110B}) 
or two-cloud modelling
\citepads{2014ApJ...792...13H}. 
On the other hand, 
\citetads{2006SoPh..235...75S}  
obtained temperatures of 1--1.2\,\dakK\ between roughly
$\ltau\tis-$3.5 and $-$5 from inversions of the \CaII\,8498\,\AA\  and 8542\,\AA\ lines
using a predecessor of the inversion code NICOLE.
A recent study by 
\citetads{2017A&A...598A..33L}, 
considering \HeIDthree\ and 10830\,\AA, suggests temperatures of order
15,000\,K would be consistent with their observations.
Profile fitting efforts by
\citetads{2016A&A...593A..32G} 
were able to reproduce the observed IRIS \MgIIh\ profiles with temperature
enhancements of \dT=1100--3350\,K (up to roughly 8000\,K total
temperature), while also yielding
\EB-like \Halpha\ profiles.

More recently,
\citetads{2017ApJ...835L..37R} 
and
\citetads{2017ApJ...845..144H} 
used the radiative hydrodynamics code RADYN 
to model \EBs\ and found that temperature enhancements of up to \dT=3000\,K yielded
\Halpha\ and \CaIR\ profiles similar to observations.
Both studies also considered \MgIIk, where a larger energy deposition was
required to obtain profile shapes similar to the observations, however,
simultaneous agreement between all three diagnostics could not be achieved in
either of those studies.
They also noted---as did
\citetads{2017RAA....17...31F}---that 
temperatures in excess of 10,000\,K caused the computed \Halpha, \CaIR\ and
continuum emission to be overestimated compared to observations.

Our results largely agree with these previous findings, yet with some notable
exceptions.
As shown in the temperature maps (\eg\ Figs.~\ref{fig:eventA_maps},
\ref{fig:eventB_maps} or \ref{fig:eventA_tmaps}), combined \CaIR\ and \CaIIK\
inversions yield total temperatures of some 7000--9000\,K close to the
temperature minimum throughout most of the events, however, localised
temperatures of 1.5\,\dakK\ are found as well.
The highest temperatures are typically associated with the blue-shifted parts
of the events (cf.~\eg\ the second and third panels of the third row in
Fig.~\ref{fig:eventA_maps} or in the first row of Fig.~\ref{fig:eventB_maps}).
Another difference with previous studies is the occurrence of marked temperature
plateaus at 2--3\,\dakK\ above $\ltau\tis-6$ for some pixels (\eg\ purple and
red samplings in the top row of Fig.~\ref{fig:eventAB_profs} or the green
sampling in the bottom row of the same figure).
However, the sensitivity of the \CaII\ (and \MgII) lines is very limited above
$\ltau\simeq-6$ (cf.~also the response functions to temperature in appendix
Fig.~\ref{fig:lte_nlte_rf}) and while in some cases
the line cores may sense (and thus require) the lower part of this temperature
rise around $\ltau\simeq-$5.5, the presence or absence of the higher-located
plateaus should not be overinterpreted.
  The case is clearly different when considering \SiIV, as both
  Figs.~\ref{fig:siiv_invprof} and \ref{fig:siiv_synthprof} suggest that
  temperature modifications at these heights may have an appreciable effect on
  the \SiIV\ line intensity.

A notable effect of taking \MgIIhk\ into consideration is that the peak
temperatures around $\ltau\tis-3$ are reduced compared to the \CaII\ inversion
results.
Intuitively, as \MgII\ is formed higher than either of the \CaII\ lines, one
would expect higher temperatures to be recovered when including the \MgIIhk\
lines (or at the very least not as strong a reduction as we find).
However, this is likely not a physical effect, but rather because of the
difference in resolution between SST/CHROMIS and IRIS (and possibly timing
differences as well), as pointed out previously in 
Section~\ref{sec:invAB_sst_iris_mg} and discussed in Section~\ref{sec:limitations}.

Both the selected pixel examples (Figs.~\ref{fig:eventAB_profs} and
\ref{fig:eventAB_mg_profs}) and the cross-cuts
(Fig.~\ref{fig:eventA_tcrossmaps}) generally show that the transition region
comes down to somewhere between $\ltau\tis-$5.5 and $-$6.
Also the temperature (difference) maps at $\ltau\tis-$4 indicate that the whole
sub-FOV is enhanced in temperature by some \dT=2000--2500\,K over the input
temperature at that height, which could be interpreted as a heating of the
chromosphere above the event.
On the other hand, recent larger field-of-view inversions by 
\citetads{2018A&A...612A..28L} 
indicate persistent and space-filling heating above $\ltau\tis-$4 during flux
emergence, wherein localised brightenings (\eg\ \EBs) only take up a fraction of
the flux emergence brightenings and may therefore play only a minor role in the
overall heating of the chromosphere.

\subsection{Reproducing \UVB\ \SiIV\ emission}
Many studies have quoted the transition region equilibrium temperature of
order 8\,\dakK\ as requirement to explain the observed \SiIV\ emission in \UVBs.
As discussed above, for those events that show \EB\ properties this
poses the obvious problems of producing too strongly enhanced \Halpha\ and
\CaII\ lines or continua.
Under certain assumptions, such as LTE at the onset of the events 
\citepads{2016A&A...590A.124R} 
or if heating were to happen in the photosphere where densities are high
\citepads{2017ApJ...839...22H}, 
lower temperatures of order 1--2\,\dakK\ may suffice to cause \SiIV\ emission,
yet this does appear to certainly be the lower limit.
Considering the localised temperatures of order 1--1.5$\times 10^{4}$\,K that we
obtained from the \CaII\ inversions, these may in fact be marginally sufficient
to cause \SiIV\ emission.
Indeed the inversions including \SiIV\ retain a temperature enhancement of that
order between $\ltau\tis-$2 and $-$4 (cf.~Fig.~\ref{fig:siiv_invprof}) and show
good fits to both \CaIR\ and \CaIIK, yet suffer notably from ill-fitted \MgII\
lines.

On the other hand, it also remains to be seen whether these temperatures are
needed as low down as \EBs\ are typically believed or inferred to occur.
From their \Bifrost\ simulations
\citetads{2017ApJ...839...22H} 
suggest that in different magnetic topology, and considering that the \EB\ jets
have a sizeable vertical extent in Joule heating, their impact may reach
sufficiently high to produce \SiIV\ emission at the \EB\ tops 
(analogous results of extended heating were found by 
\citetads{2017ApJS..229....5D}  
for weaker-field \EB-like events in MURaM simulations).
One could envision this as shock heating at the pertinent heights, but a similar
effect could be reached if reconnection cascades further upwards, in which case
the actual reconnection at $\sim$2\,Mm heights would result in the typical \UVB\
emission.
Such offsets between the main \Halpha\ and \SiIV\ emission are not evident
in our data (likely because the viewing angle is too close to the vertical
in both data sets), but small offsets were already reported by
\citetads{2015ApJ...812...11V} 
and should be more easily disentangled closer to the limb. 
On the other hand, this should not be an issue {\it a priori} for the inversions
with STiC, as it could adapt the temperature profile to such a scenario,
provided sufficient depth resolution in the temperature node specification.

The purple profiles in Fig.~\ref{fig:siiv_invprof} are a clear example of such
a case with two separate heating sources, the lower one around $\ltau\tis-$3
explaining the \CaII\ profiles, while the upper one around $\ltau\tis-$6 is
likely responsible for the \SiIV\ emission (inspection of its response
function to temperature perturbations shows maximum response between
$\ltau\tis-6$ and $-$6.5 around the \SiIV\,\,1394\,\AA\ and 1403\,\AA\ rest wavelengths).
In fact, comparing with the other two samplings, for that case both \CaII\ lines
and \SiIV\ show better fits and also the \MgII\,\kthree\ core and both \ktwo\
peaks are well-reproduced in absolute intensity (and even shape, out to about
$\pm$0.25\,\AA).
Even though it does not resolve the general overestimation of \MgII, considering
the synthesis results of high- versus low-atmosphere temperature peaks, this may
indeed be a direction in which to seek the solution. 
In addition, while the offset between the temperature peak locations appears
large in \ltau, in terms of physical height this may in fact be squeezed closer
together in the presence of strong magnetic fields---not a far-fetched
assumption under these circumstances.

The synthesis results (Section~\ref{sec:siiv_spikes}) indicate that
the the addition of sharp temperature enhancements above $\ltau\simeq-$6 would
not strongly affect the \CaII\ and \MgII\ while providing sufficient \SiIV\
emission to at least reach the observed intensities (though not broadening).
At heights below about $\ltau\tis-$5 even the sharpest enhancements
considered (of base width $\ltau\tis0.05$) with sufficient temperature to yield
\SiIV\ emission are incompatible with the \MgII\ profiles.
Implementing ``temperature spike''-fitting capibility may be worthwile to
explore further, even though likely not sufficient in and of itself to attain
agreement with all observables.
All in all, the \SiIV\ results indicate that if its emission indeed
primarily originates around $\ltau\tis-$6, temperatures of order 3.5--6.0\,\dakK\
could suffice, rather than the often-quoted 8\,\dakK.

We also investigated whether the assumption of LTE versus non-LTE electron
densities could have made a difference. 
The comparison in Appendix~\ref{sec:appendix} shows that while the overall
temperature and velocity stratification are similar for both inversions
combining \CaIIK\ and \CaIR, and those including \MgII\ as well, individual
pixels may show more pronounced effects.
Moreover, Fig.~\ref{fig:lte_nlte_nne} suggests larger deviations between the LTE
and non-LTE electron densities above \ltau$\simeq-$4 (with typically lower
values for the latter), a height above which the temperature profiles in
Fig.~\ref{fig:siiv_invprof} also show the largest change with respect to the
inversions without \SiIV. 
Hence, the results with \SiIV\ may be more strongly affected by the choice for
LTE or non-LTE electron density.
Unfortunately, stability of the inversions was a limiting issue here and while
we attempted different approaches (including use of a more extended 23-level
\SiIV\ model atom) we could converge atom populations for only two samplings
(blue and purple), in both cases reproducing the \SiIV\ peak intensities but not
their asymmetries, and failing altogether for the third (red) sampling.
However, when running our synthesis tests assuming non-LTE electron densities,
the sharp temperature enhancements had to be placed at smaller \ltau\ values
(\ie\ typically higher electron densities) to achieve similar response in the
\SiIV\ lines, suggesting that in the inversions the temperature enhancement may
move to somewhat lower heights compared to the results presented in
Fig.~\ref{fig:siiv_invprof}.

Alternatively, the solution should perhaps not be sought thermally alone.
Another way of producing \SiIV\ without the need for excessively high
temperatures in the lower atmosphere is through high-energy particles 
produced during the reconnection.
For instance, 
\citetads{2017A&A...603A..14D}  
showed that the ionisation rates increase dramatically at low temperatures in
the presence of accelerated particles (even if they only make up a small
fraction) and may extend the formation temperatures of \SiIV\ down to
1--1.5\,\dakK. 
Our \CaII\ (and \SiIV) inversions yield very similar temperatures in localised
hot pockets and may thus be compatible with this picture.

\subsection{Line-of-sight velocity patterns}
In the reconnection scenario, outflows from the reconnection point---\ie\
bi-directional jets---are to be expected and this has been corroborated for both
\EBs\ and \UVBs\ from previous observations and numerical experiments. 
Of the events we investigated, Event A shows the clearest spatially separated
bi-directional jet signature with line-of-sight velocities of order 15-25\,\kms\
both towards and away from the observer, though with slightly higher blue
shifts.
The time evolution of this event (Fig.~\ref{fig:eventA_tmaps}) shows this is a
persistent signature with similar velocities throughout the event's lifetime.
\EB\ velocities quoted in the past are typically much lower, a few \kms\ at most
both from observations
(\eg\
\citeads{2008PASJ...60...95M}) 
and simulations
\citepads{2009A&A...508.1469A}, 
with somewhat higher values from inversions
(\eg\
\citeads{2006SoPh..235...75S};  
\citeads{2017A&A...598A..33L}),  
however our highest values are similar to those obtained from the \Bifrost\
numerical experiments of \EBs\ by
\citetads{2017ApJ...839...22H}. 
When including \MgII\ (and in particular due to the contribution from the \MgII\
triplet lines) the red-shifts are more concentrated at lower heights and not as
pronounced above $\ltau\tis-$3 as they were when considering \CaII\ only
(cf.~Figs.~\ref{fig:eventA_maps} and \ref{fig:eventA_mg_maps}).

Event B shows somewhat lower velocities, in particular in its second (B2)
snapshot.
The first snapshot exhibits the blob-like substructure that 
\citetads{2017ApJ...851L...6R} 
interpreted as plasmoids, an important ingredient in their argument that the
broadening and non-Gaussian shapes commonly observed in \UVB\ \SiIV\ spectra may
result, at least in part, from a superposition of plasmoid blobs of different
line-of-sight velocities within the IRIS resolution element.
Assuming that the \SiIV\ emission would originate in the same structures, the
line-of-sight velocities inferred from \CaII\ are insufficient to
explain the broadening observed in \SiIV.
However, if the latter emission would originate in plasmoid-connected shocks
(Ni et al.~\citeyearads{2018ApJ...852...95N}, 
\citeyearads{2018arXiv180405631N}) 
these may in fact attain sufficiently high velocities to explain broadening at
least out to some 50\,\kms.

\subsection{Magnetic fields}
One of our aims was to characterise the magnetic field configuration of \EBs\
with \UVB\ signature, but due to the limitations of the data (\ie\ no
well-constrained photospheric fields for September 3 and in general limited 
chromospheric field sensitivity due to observing program choices),
reconstructing the topology through the atmosphere is a challanging task
at best.  
We do, however, find increased horizontal fields co-spatial with the stronger
intensity enhancements (cf.~Figs.~\ref{fig:eventB_maps} and
\ref{fig:eventB_Bmaps})
which is consistent with the $\cup$-loop reconnection scenario suggested for
both \EBs\ and \UVBs\ in many observational studies (\eg\
\citeads{2002ApJ...575..506G}, 
\citeads{2004ApJ...614.1099P}, 
\citeads{2008PASJ...60..577M}, 
\citeads{2009ApJ...701.1911P}, 
\citeads{2010PASJ...62..879H}) 
and established in a number of numerical studies as well (\eg\
\citeads{2009A&A...508.1469A},  
\citeads{2017A&A...601A.122D}, 
\citeads{2017ApJ...839...22H}). 
Furthermore, the changes from snapshot B1 to B2 at 1.5\,min interval suggest the
horizontal fields decrease and become near invisible at lower heights, while
retaining some signal higher up and at the same time an opposite-polarity
signature remains visible in the line-of-sight component.
This could be interpreted as seeing the imprint of a $\cap$-loop topology or
possibly at the post-reconnection $\cup$-shaped fields rising further through
the atmosphere while the $\cap$-shaped fields below the reconnection point sink
further down.
However, observations with higher polarimetric sensitivity are required to
better constrain the inferral of (low-)chromospheric fields and their evolution.

\subsection{Limitations of the inversion approach}
\label{sec:limitations}
We have obtained results that are consistent with the observations (in
terms of intensities and profile asymmetries) and theoretical and numerical
studies (\eg\ the presence of bi-directional flow signatures, temperature
enhancements close to the temperature minimum, enhanced temperatures at
locations of enhanced horizontal fields, etc.), in particular when
combining CRISP and CHROMIS observations. 
However, when including IRIS diagnostics we are unable to fit
all observables simultaneously with the proposed models. 
Particularly challenging appears to be the reconciliation of \MgII\ with the
other diagnostics in the presence of \SiIV.
This suggests we reached certain limitations of our approach, which we have
already largely discussed before.
We shortly summarise them here:
\begin{enumerate}
  \item The spectra are not strictly co-temporal, even though STiC works under
    the assumption that they are. 
    For the cases presented this can range anywhere between 2.3--9.2\,s, meaning
    the effects on following fast-evolving substructure can be substantial.
    While this could in principle be minimised further, this is not always
    feasible as seeing effects also play a role.
  \item The instrumental resolution differences are large, in
    particular between CHROMIS and IRIS.
    The choice not to sacrifice high-resolution means a single IRIS spectral
    profile is spread over many SST pixels and thus compared with profiles that
    are not strictly co-spatial at the CHROMIS pixel level.
    Test inversions of combined SST and IRIS observations, where SST data were
    downsampled to IRIS resolution (so as to mimick taking the resolution
    differences into account), yielded similar results as the high-resolution
    inversions, but with typically lower $\chi^{2}$ values for the profile fits
    (and while to a lesser extent, \MgII\ remained overestimated in the presence
    of \SiIV). 
    This suggests that proper accounting for the resolution difference would
    indeed improve results, but may not be sufficient in itself to reconcile all
    observables.
  \item While 3D radiative transfer effects are important for the \CaII\ and
    \MgII\ line cores, these do likely not play a determining role in the \SiIV\
    formation.
    Nonetheless, the pixel-by-pixel inversion may be too restrictive, \eg\ if the
    \SiIV\ emission were to originate from the low-altitude temperature enhancement,
    radiation has no means to escape sideways and will heat up the entire pixel
    atmosphere, which may explain part of the \MgII\ overestimation found.
    On the other hand, the densities are much higher close to the temperature
    minimum than in the upper chromosphere and the effects of horizontal
    scattering therefore expectedly smaller.
  \item The node-based inversions are limited in resolving what in reality must
    be continuous atmosphere.
    While we did not observe strong effects on the line-of-sight velocity this
    may play an important role for the temperature stratification, in particular
    if \SiIV\ originates in temperature enhancements that are very localised
    with height.
\end{enumerate}

Spatially-coupled (\ie\ two-dimensional) inversions could provide a
solution to---or at the very least alleviate---some of these problems (notably
(3) and likely also (2) in the list above). 
The idea behind this is that the spectra in adjacent pixels are not independent,
both due to observational effects (\eg\ smearing by the telescope point spread
function (PSF)) and physical ones (\eg\ horizontal radiative transfer), and an
evident improvement over the 1.5D inversions performed here would thus be to
account for this (horizontal) spatial coupling.
  A promising two-dimensional inversion approach is the one proposed by 
\citetads{2015A&A...577A.140A} 
  and builds on the concept of sparsity, providing dual gains as it not only
  reduces the number of unknowns, but also ensures spatial coupling given that a
  reduced fraction of elements describes the behaviour of all pixels.
In addition, taking the PSFs of the different instruments into account 
(similar to the approach for \Hinode\ alone in 
\citetads{2012A&A...548A...5V}) 
would solve the previously stated resolution difference issues.
Unfortunately, at this point we are not able to explore this further, since the
current STiC code design does not allow for implementation of such a
two-dimensional approach.

Finally, the assumption of hydrostatic equilbrium is a counterintuitive one
considering the dynamics of \EBs\ and \UVBs, but since the quantities are
derived with respect to column mass rather than in a physical height scale,
this is in fact not an unreasonable simplification for radiative transfer
calculations.
We do note that hydrostatic equilibrium prescribes a monotonic increase of gas
pressure inward, meaning that bumps and discontinuities in pressure (\eg\ as a
result of (bi-directional) jets or shocks) are impossible to reproduce with this
code and may in turn lead to locally over/underestimated temperatures and
densities.
However, we believe other factors play a larger role and that the uncertainties
are primarily set by the limited number of nodes, the absence of spatial
coupling between the solutions and the effects of greatly different instrumental
resolution.

\section{Conclusion}  \label{sec:conclusion}
We have presented first-time non-LTE inversions of \CaIR, \CaIIK, \MgII\ and
\SiIV\ in \EBs\ with \UVB\ signatures, using the STockholm Inversion Code---a
powerful tool allowing multi-line, multi-species non-LTE inversions.
The revived interest in \EBs\ over the past decade-and-a-half has led to a
better understanding of the phenomenon, but also added to the diagnostic
visibilities that require explanation, in particular since the launch of IRIS.
As such STiC is particularly well-suited to address the pressing issue of
explaining this wide range in diagnostic formation temperatures from a seemingly
limited atmospheric volume.

We have found that we can largely reproduce the observational properties of the
events (\eg\ specific intensities, profile asymmetries and morphology) with
temperature stratifications that typically peak close to the classical
temperature minimum and velocity profiles that suggest bi-directional jet flows.
The inferred temperatures fall partly in the range of earlier expections with
enhancement of a few thousand kelvin, yet we also find localised hot pockets of
up to 15,000\,K when considering SST diagnostics only.
In general the atmospheric parameters are better constrained when including more
diagnostics and the addition of \MgII\ appears to yield more moderate
temperature enhancements, while the \MgII\ triplet lines help constrain the
low-atmosphere velocity gradients.
The assumption of LTE versus non-LTE hydrogen ionisation appears to have
  little effect on the spatial distribution of the event heating, both in the
  observed plane and in \ltau\ height, but likely plays a more prominent
role for \SiIV.
  The latter's intensities can be reproduced in double-peaked temperature
  stratifications with enhancements of 35,000--60,000\,K around
  $\ltau\tis-$6, while requiring in excess of 10,000\,K if the \SiIV\ emission
  should originate close to the temperature minimum.

At the same time it is also clear that we run into certain limitations of our
approach, as with the current setup and inferred model atmospheres we are unable
to reproduce all \UVB\ and \EB\ signatures in full agreement simultaneously. 
This is likely a combined effect of the difference in instrument resolution,
non-zero time difference between the acquisition of the spectra (which given the
fast evolution of the substructure may represent a significant effect) and also
the limitation to pixel-by-pixel plane-parallel atmospheres.
For the case of \SiIV\ emission, our study suggests that considering
double-peaked temperature solutions and allowing sharp temperature enhancements
may be worth exploring further. 
Ultimately, though, dealing differently with the instrument pixel size
differences---including
moving to spatially-coupled inversions---is
likely a necessary step to reach better diagnostic agreement and by extension a
more complete picture.

\begin{acknowledgements}
This work was supported under the CHROMOBS grant by the Knut and Alice
Wallenberg Foundation. 
JdlCR is supported by grants from the Swedish Research Council (2015-03994), the
Swedish National Space Board (128/15) and the Swedish Civil Contingencies Agency
(MSB). This project has received funding from the European Research Council
(ERC) under the European Union's Horizon 2020 research and innovation programme
(SUNMAG, grant agreement 759548).
This research is also supported by the Research Council of Norway, project number
250810, and through its Centres of Excellence scheme, project number 262622.
The Swedish 1-m Solar Telescope is operated on the island of La Palma by the
Institute for Solar Physics of Stockholm University in the Spanish Observatorio
del Roque de los Muchachos of the Instituto de Astrof\'isica de Canarias. 
IRIS is a NASA small explorer mission developed and operated by LMSAL with
mission operations executed at NASA Ames Research center and major contributions
to downlink communications funded by ESA and the Norwegian Space Centre.
We are grateful to Shahin Jafarzadeh, Tomas Hillberg and Pit S\"utterlin for
participating in the observations at the SST and to Paul Bryans as IRIS planner
for the IRIS--SST coordination.
The inversions were performed on resources provided by the Swedish National
Infrastructure for Computing (SNIC) at the High Performance Computing Center
North at Ume\aa\ University.
This work profited from discussions at the meetings ``Solar UV bursts -- a new
  insight to magnetic reconnection'' (International Team 360) and ``Studying
magnetic-field-regulated heating in the solar chromosphere'' (International Team
399) at the International Space Science Institute (ISSI) in Bern.
We made much use of NASA's Astrophysics Data System Bibliographic Services.
We also acknowledge the community effort to develop open-source
packages used here: \tt{numpy} (\url{numpy.org}), \tt{matplotlib}
(\url{matplotlib.org}), \tt{sunpy} (\url{sunpy.org}).

\end{acknowledgements}

\bibliographystyle{aa}

\bibliography{journals,rjrfiles,adsfiles,chromis-uvbinv} 
%

\appendix
\section{The effect of LTE versus non-LTE hydrogen ionisation on the inversion
results}\label{sec:appendix}
With hydrogen ionisation being the primary source of electrons in the solar
atmosphere, the assumption of either LTE or non-LTE ionisation has a direct
bearing on the derived electron densities, in turn affecting the emergent line
profiles because of the degree of collisional coupling between the source
function and the local conditions. 
From the inversion point of view, this means that a fit to a spectral profile
could lead to a different temperature stratification given a different electron
density stratification.
STiC includes the option to derive electron densities directly from an LTE
equation of state or make them consistent with non-LTE hydrogen ionisation, by
iteratively solving the statistical equilibrium equations while imposing charge
conservation.
In this appendix we assess the impact on our results of considering either case. 

\begin{figure*}[htp]
  \centerline{\includegraphics[width=\textwidth]{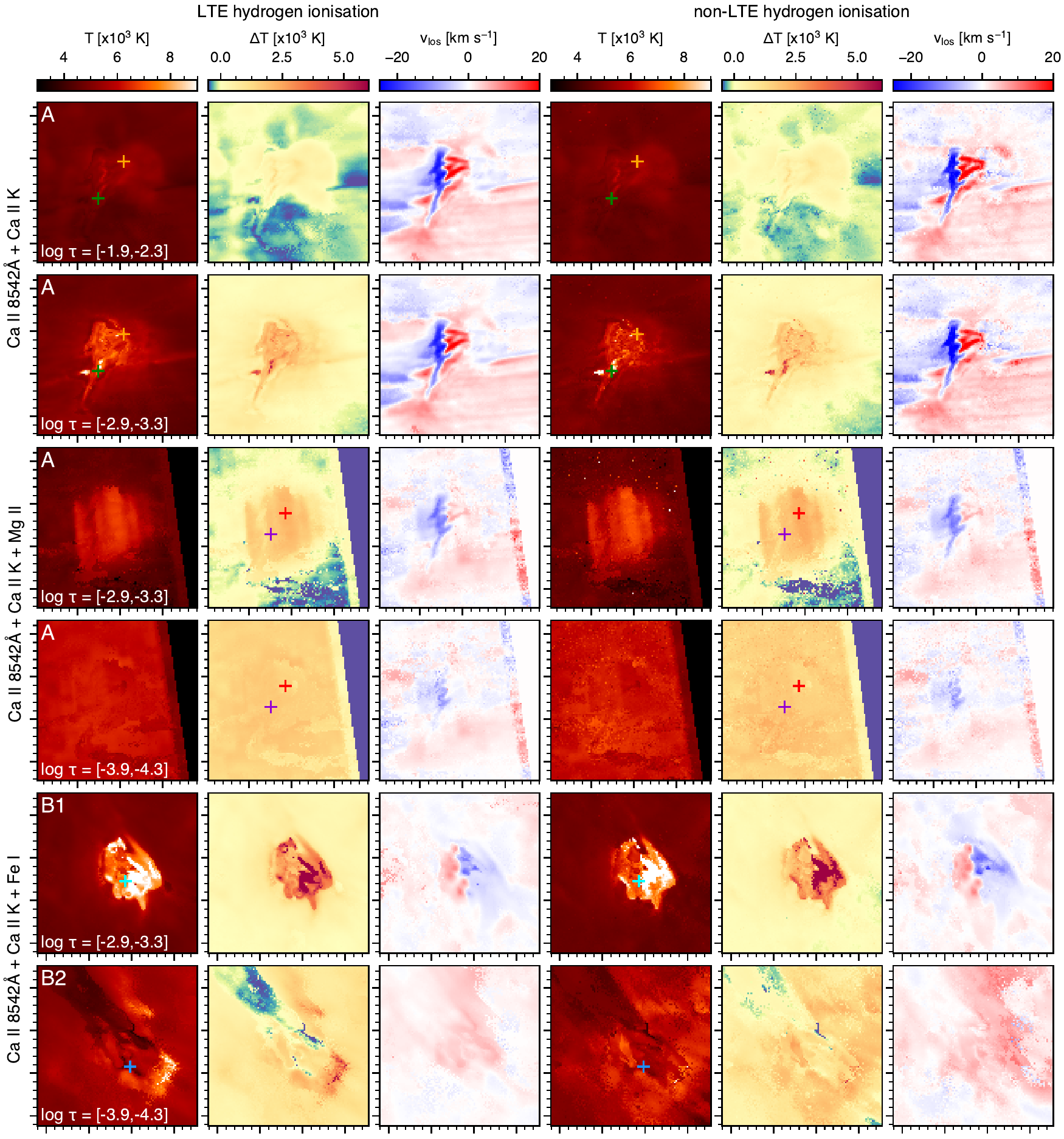}}
  \vspace{-1.5ex}
  \caption[]{\label{fig:lte_nlte_maps} %
    Inversion maps assuming hydrogen ionisation in LTE ({\it left-hand three
    columns\/}) and in non-LTE ({\it right-hand three columns\/}).
    Each set of columns has the same format as the left-hand three columns of
    \eg\ Fig.~\ref{fig:eventA_maps}, \ie\ showing from left to right the
    temperature, the temperature difference to the input FAL-C temperature and
    the line-of-sight velocity.
    The maps correspond to inversions combining \CaIR\ and \CaIIK\ for event A
    ({\it top two rows\/}), \CaII\ and \MgII\ for the same event ({\it middle
    two rows\/}), and \CaII\ and \FeI\ for event B ({\it bottom two rows\/}).
    Coloured markers correspond to pixels for which detailed profiles are
    shown in Fig.~\ref{fig:lte_nlte_profs} and are overlaid in the first and
    third columns (except for the middle two rows where they have been overlaid
    in the second and fourth column panels instead for better visibility). 
    These sampling locations are the same as the identically coloured locations
    in Figs.~\ref{fig:eventA_maps} ({\it top two rows\/}),
    \ref{fig:eventA_mg_maps} ({\it middle two rows\/}) and \ref{fig:eventB_maps}
    ({\it bottom two rows\/}).
    Major tick marks are spaced 1\arcsec\ apart for all panels.
  }\vspace{-2ex}
\end{figure*}

Figure~\ref{fig:lte_nlte_maps} compares inversion maps at selected heights for
all events presented in this publication.
The spatial distribution of temperatures
and line-of-sight velocities within our features of interest are essentially the
same, regardless of the LTE or non-LTE ionisation choice.
Here the most significant differences are found at $\ltau\tis-$4 for the second
snapshot of Event B (last row), where the elongated cold feature intruding from
the top left of the panel seems to be at higher temperatures for the non-LTE
ionisation case (\ie\ it practically disappears in the temperature difference
panel).
Similarly, the cold region at $\ltau\tis-$3 below Event A (top row) is also
warmer in the inversions with non-LTE electron densities.
On the other hand, in the inversion including \MgII\ (third and fourth rows)
this cold region at $\ltau\tis-$3 is in fact more extended, while the
temperature maps at $\ltau\tis-$4 appear practically identical.
The non-LTE ionisation results are also more noisy, which is indicative of
pixels where convergence failed and therefore of less stable inversions overall.
This is particularly noticeable in the temperature (difference) maps in the
third row (Event A with \MgII), but also when comparing the \vlos\ panels in the
top two (Event A) and the bottom (Event B2) rows.

\begin{figure*}[ht]
  \centerline{\includegraphics[width=\textwidth]{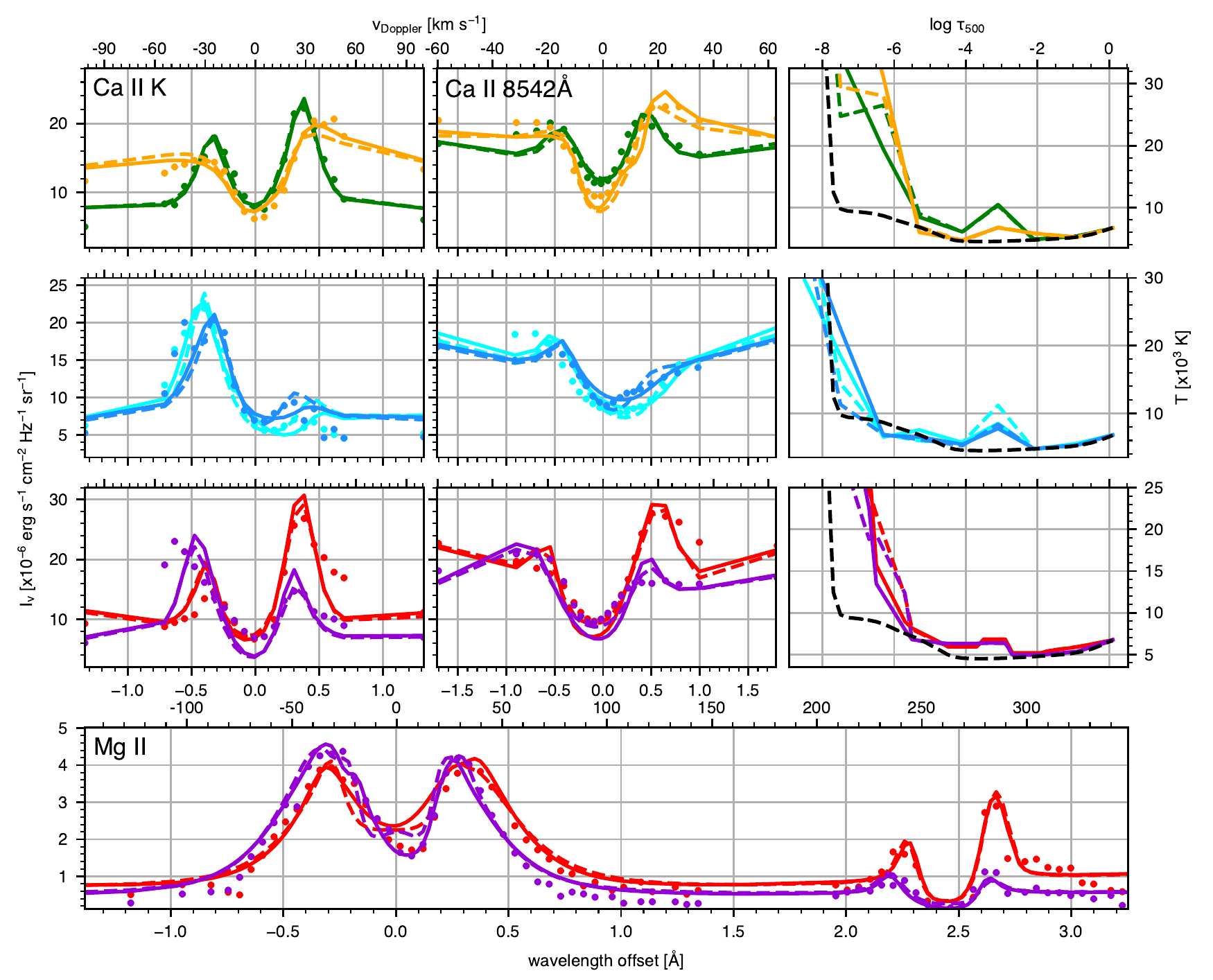}}
  \vspace{-2ex}
  \caption[]{\label{fig:lte_nlte_profs} %
    Spectral line and temperature profiles for six selected pixels in Events A
    ({\it top and bottom two rows\/}) and B ({\it second row\/}) at the
    identically coloured locations marked in Fig.~\ref{fig:lte_nlte_maps}.
    The left-hand, middle and bottom panels show fits assuming hydrogen
    ionisation in LTE ({\it dashed lines\/}) and non-LTE ({\it solid lines\/}) to
    the observed profiles ({\it filled circles\/}) for \CaIIK\ ({\it left panel
    in the first three rows\/}), \CaIR\ ({\it middle panel in the same rows\/})
    and \MgII\ ({\it bottom panel\/}).
    The corresponding temperature stratification ({\it right-hand panels\/}) is shown
    using the same colour coding, along with the FAL-C input temperature for
    reference ({\it dashed black line\/}).
  }
\end{figure*}

Figure~\ref{fig:lte_nlte_profs} highlights the changes in spectral profile fits
and temperature stratification between the two approaches for selected pixels in
Fig.~\ref{fig:lte_nlte_maps}. 
The differences between the LTE case (dashed lines) and the non-LTE case (solid
lines) are generally small in the profile fits, with sometimes better results
when assuming LTE densities (\eg\ the red-wing \CaIR\ peak for the orange
sampling in the top row, the purple \CaII\ \KtwoR\ peak in the third row, or the
red \CaIR\ profile in the same row), while for others the assumption of non-LTE
electron densities yields better results (\eg\ the orange \CaIIK\ profile or the
core of the purple \MgII\ sampling in the bottom panel).
For again others the results are practically identical (\eg\ the green
samplings in the top row, the red \CaIIK\ one in the third row, or the fits to
the \MgII\ triplet lines at $\Delta \lambda = 2.5$\,\AA\ in general).
Overall, the largest differences appear when including \MgII\ in the inversions,
suggesting that the choice for LTE or non-LTE ionisation has a bigger impact
there.

The temperature stratifications exhibit more significant changes, though mostly
above $\ltau\tis-$5 to $-$6 (where the sensitivity of the spectral lines is
minimal). 
The localised temperature increase around $\ltau\tis-$3 is practically unchanged
in magnitude for all except the cyan sampling.
When assuming non-LTE electron densities, the chromospheric temperature plateaus
around 2.5--3.0\,\dakK\ for the \CaII\ inversions in the top row disappear and for
Event B (second row) the transition region temperature rise is moved to slightly
lower heights, while for the inversions with \MgII\ the opposite is the case in
exchange for a steeper temperature increase.
However, given the highly reduced sensitivity of both \CaII\ and \MgII\ lines
above $\ltau\tis-$6 (as also evidenced by the response functions to temperature
discussed further down) the presence or absence of the temperature
plateaus at those heights is not well-constrained by the spectral profiles and
should thus be interpreted with great care. 

\begin{figure*}[ht]
  \centerline{\includegraphics[width=\textwidth]{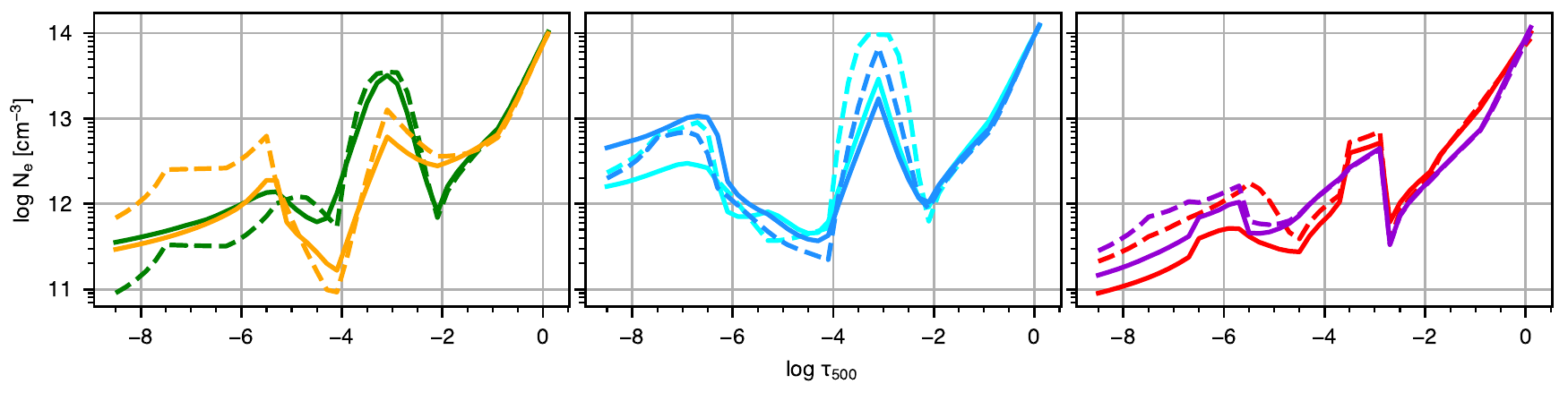}}
  \vspace{-2ex}
  \caption[]{\label{fig:lte_nlte_nne} %
    Logarithm of electron density $N_{\rm e}$ as function of \ltau\ in
    the selected pixels from Figs.~\ref{fig:lte_nlte_maps} and
    \ref{fig:lte_nlte_profs}, with identical colour coding.
    The curves display the densities consistent with either LTE ({\it dashed\/}) or
    non-LTE ({\it solid\/}) hydrogen ionisation.
  }
\end{figure*}

Notwithstanding, these changes are not surprising considering the differences in
the electron density stratification depending on the LTE or non-LTE ionisation
assumption.
Figure~\ref{fig:lte_nlte_nne} presents these densities as function of \ltau\ for
all samplings shown in Fig.~\ref{fig:lte_nlte_profs}.
As one would expect for the lower atmosphere, the differences below $\ltau\tis-$2
to $-$2.5 are negligible for all samplings and where those from Event A
(green, orange and red) exhibit only minor offsets between the LTE and non-LTE
electron densities around the temperature peak at $\ltau\tis-$3, the deviations
are more promininent for Event B (blue and cyan in the middle panel).
This is consistent with the similarity in the temperature enhancement in the case
of LTE versus non-LTE (Fig.~\ref{fig:lte_nlte_profs}, last column) around
$\ltau\tis-$3 for the Event A samplings, while those for Event B are visibly
different (by over 2600\,K for the cyan sampling). 
Above $\ltau\tis-$4 to $-$5 the LTE and non-LTE densities are disparate for all
pixels, in most cases also showing the largest deviations, which again coincides
with a marked change in temperature stratification above those heights.

\begin{figure*}[bt]
  \centerline{\includegraphics[width=\textwidth]{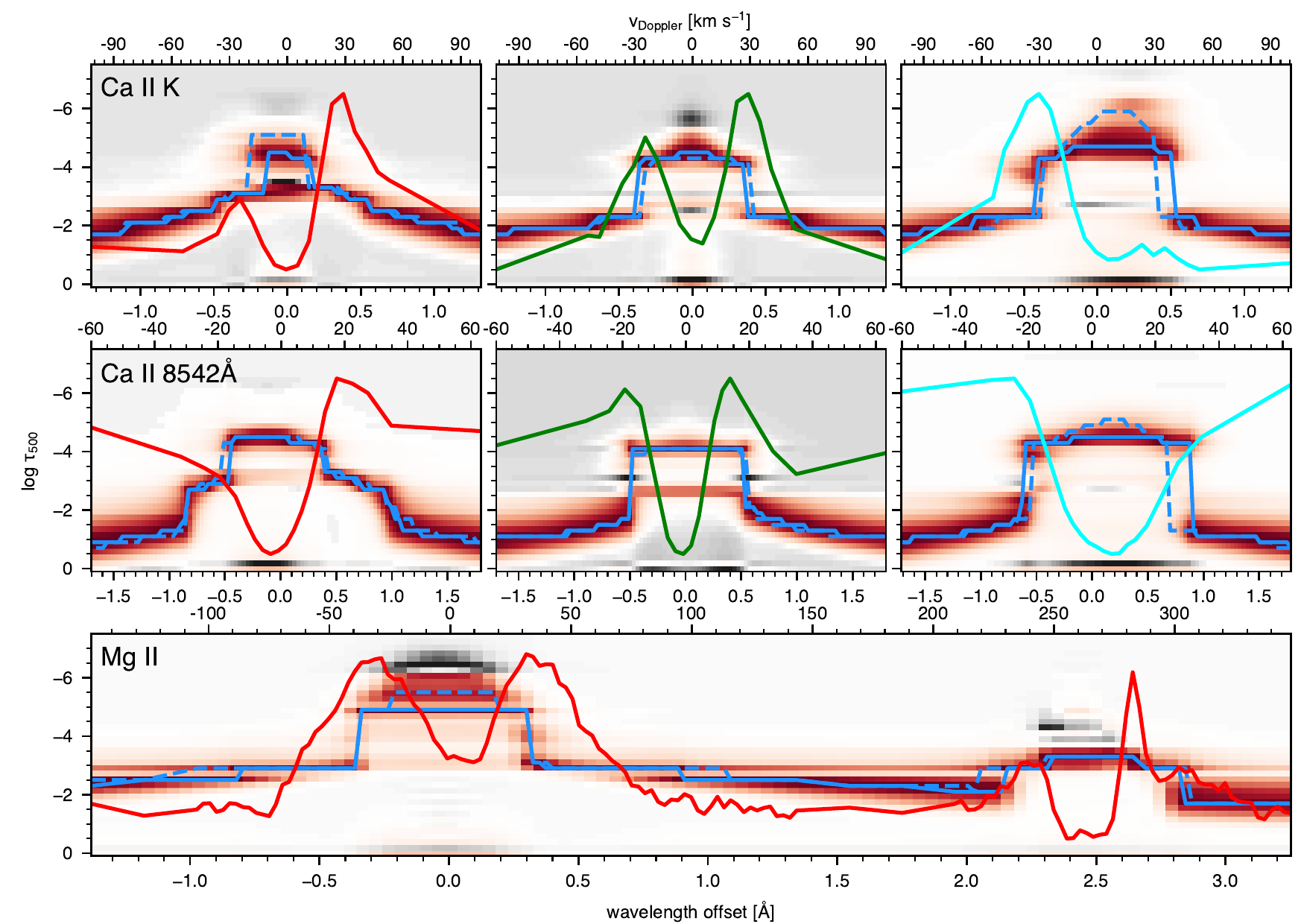}}
  \caption[]{\label{fig:lte_nlte_rf} %
    Response functions to temperature with \ltau\ as function of wavelength in
    \CaIIK\ ({\it top row\/}), \CaIR\ ({\it middle row\/}) and \MgII\ ({\it
    bottom panel\/}) for three of the selected pixels from
    Figs.~\ref{fig:lte_nlte_maps} and \ref{fig:lte_nlte_profs}.
    Results are shown for the red sampling ({\it first column and bottom
    panel\/}) and green sampling ({\it second column\/}) of event A, and the
    cyan sampling of event B ({\it third column\/}), under the assumption of
    non-LTE hydrogen ionisation.
    Corresponding observed \CaII\ and \MgII\ profiles are shown for reference
    and have been plotted on arbitrary scale to maximise visiblity of profile
    features.
    In each panel the response function is displayed in red-white-black colour
    scale (with red being positive response, black negative and white none), at
    each wavelength normalised to its maximum with height \ltau. 
    The blue lines indicate the \ltau\ heights as function of wavelength for
    which the response function peaks in the case of LTE ({\it dashed\/}) and
    non-LTE ({\it solid\/}) electron densities.
  } 
\end{figure*}

Finally, Fig.~\ref{fig:lte_nlte_rf} presents the wavelength-dependent response
functions to temperature perturbations for three of the previously considered
samplings (red, green and cyan), with positive response in red and negative
response in black.
These show that in the line wings where the events are best viewed (\ie\
the peaks at $\pm$30\,\kms\ for \CaIIK, $\pm$20\,\kms\ for \CaIR, and out at
$\pm$(40--50)\,\kms\ for \MgIIk) the peak response (blue lines) covers a range from
$\ltau\tis-$2.5 to $-$4 for \CaII\ and up to $-$5 for \MgII, irrespective of the
electron density choice and coinciding with the lower-atmospheric temperature
enhancements as shown in Fig.~\ref{fig:lte_nlte_profs}.
In addition, at the red wing enhancement of the red sampling (left hand and
bottom panels) both the \CaII, \MgIIk\ and the triplet lines exhibit the
strongest response to temperature at similar \ltau\ heights.
The choice of LTE versus non-LTE electron densities primarily affects the line
cores, being sensitive to temperature perturbations at
$\ltau\tis0.5-1.0$ higher heights (\eg\ the red \CaIIK\ and \MgIIk\ or the cyan
\CaII\ samplings) in the LTE case, while there is little to no effect in the
line wings.
This may also explain why the largest differences are observed in the \MgIIk\
line core, as its response to temperature perturbations reaches to higher
heights than \CaII.

Considering these results, our bottom line conclusion is therefore that when
inverting \CaII\ and \MgII\ for \EBs\ with \UVB\ signature the assumption of LTE
hydrogen ionisation is adequate as it leads to essentially the same results as
when assuming non-LTE hydrogen ionisation.
As \SiIV\ is also sensitive to temperature perturbations at higher heights,
electron densities consistent with non-LTE ionisation may be required for a
correct description of the problem in that case.

\end{document}